\newif\ifnotes

\documentclass[runningheads]{llncs}

\usepackage{graphicx}
\usepackage{hyperref,doi}
\usepackage[utf8]{inputenc} %
\usepackage[T1]{fontenc} %

\usepackage[inference]{semantic}
\usepackage[obeyFinal]{todonotes}
\usepackage{bm}
\usepackage{amssymb}
\usepackage{amsthm}
\usepackage{mathtools}
\usepackage{xspace}
\usepackage{stmaryrd}

\usepackage{algorithm}
\usepackage{algpseudocode}%
\usepackage[usenames,dvipsnames,table]{xcolor}
\usepackage{proof}
\usepackage{fontawesome5}

\usepackage[
  sorting=none,
  style=numeric-comp,
  backend=biber,
  isbn=false,
  url=true,
  doi=false,
  url=false,
  mincrossrefs=100,
  maxnames=12,
  firstinits=true
]{biblatex}

\ifnotes
\usepackage[nomargin,inline,draft]{fixme}
\else
\usepackage[nomargin,inline,final]{fixme}
\fi

\usepackage{pifont}

\makeatletter
\def\footnotesize{%
  \@setfontsize\footnotesize\@viiipt{9.5}%
  \abovedisplayskip 6\p@ \@plus2\p@ \@minus4\p@
  \abovedisplayshortskip \z@ \@plus\p@
  \belowdisplayshortskip 3\p@ \@plus\p@ \@minus2\p@
  \def\@listi{\leftmargin\leftmargini
    \topsep 3\p@ \@plus\p@ \@minus\p@
    \parsep 2\p@ \@plus\p@ \@minus\p@
    \itemsep \parsep}%
  \belowdisplayskip \abovedisplayskip
}%
\makeatother

\hypersetup{
  colorlinks=true,
  linkcolor=blue,
  citecolor=red
}

\AtEveryBibitem{%
  \clearname{editor}%
}
\AtEveryBibitem{%
  \clearfield{note}%
}
\AtEveryBibitem{%
  \clearlist{language}%
}

\algblockdefx{Event}{EndEvent}[1]%
{\textbf{upon event }#1 \textbf{do }}%
{\textbf{}}

\algblockdefx{Repeat}{EndRepeat}[1]%
{\textbf{repeat }#1 \textbf{times }}%
{\textbf{end repeat}}

\algblockdefx{Function}{EndFunction}[1]%
{\textbf{Function }#1 }%
{\textbf{}}

\algnewcommand\algorithmicswitch{\textbf{switch}}
\algnewcommand\algorithmiccase{\textbf{case}}
\algnewcommand\algorithmicassert{\texttt{assert}}
\algnewcommand\Assert[1]{\State \algorithmicassert(#1)}%
\algdef{SE}[SWITCH]{Switch}{EndSwitch}[1]{\algorithmicswitch\ #1\ \algorithmicdo}{\algorithmicend\ \algorithmicswitch}%
\algdef{SE}[CASE]{Case}{EndCase}[1]{\algorithmiccase\ #1}{\algorithmicend\ \algorithmiccase}%
\algtext*{EndSwitch}%
\algtext*{EndCase}%

\definecolor{vrpink}{RGB}{255,0,127}
\definecolor{vrlpink}{RGB}{255,192,203}

\definecolor{jyellow}{RGB}{255,170,29}
\definecolor{jemyellow}{RGB}{255,244,79}

\FXRegisterAuthor{vv}{avr}{\color{vrpink}[vince]\normalfont}

\FXRegisterAuthor{sls}{asls}{\color{LimeGreen}[Sam]\normalfont}

\FXRegisterAuthor{jem}{ajem}{\color{jyellow}[Javier]\normalfont}

\newcommand{\hide}[1]{}

\newcommand{\LtyAgent}[0]{\textbf{\textsc{Agent}}}
\newcommand{\LtyAgents}[0]{\textbf{\textsc{Agents}}}

\newcommand{\LtyData}[0]{\textbf{\textsc{Data}}}
\newcommand{\Echo}[1]{\mathtt{Echo}(#1)}
\newcommand{\Deliver}[1]{\mathtt{Deliver}(#1)}

\newcommand{\broadcast}[1]{{\tt Bcast}(#1)}
\newcommand{\electedSYMB}{{\tt Elected}}
\newcommand{\elected}[1]{\electedSYMB(#1)}
\newcommand{\election}[1]{{\tt Propose}(#1)}

\newcommand{\nhbrSYMB}{{\tt nhbr}}
\newcommand{\nhbr}[1]{\nhbrSYMB(#1)}

\newcommand{\OPtrue}{\top}
\newcommand{\OPfalse}{\bot}
\newcommand{\OPnegSYMB}{\neg}
\newcommand{\OPneg}[1]{\OPnegSYMB#1}
\newcommand{\OPsinceSYMB}{{\bf S}}
\newcommand{\OPsince}[2]{#1\,\OPsinceSYMB\,#2}
\newcommand{\OPuntilSYMB}{{\bf U}}
\newcommand{\OPuntil}[2]{#1\,\OPuntilSYMB\,#2}
\newcommand{\OPnextSYMB}{\bigcirc}
\newcommand{\OPnext}[1]{\OPnextSYMB\,#1}
\newcommand{\OPprevSYMB}{{\bf Y}}
\newcommand{\OPprev}[1]{\OPprevSYMB\,#1}
\newcommand{\OPprevbSYMB}{{\bf Y}_{?}}
\newcommand{\OPprevb}[1]{\OPprevbSYMB\,#1}
\newcommand{\OPopSYMB}{\sim}
\newcommand{\OPop}[2]{#1\,\OPopSYMB\,#2}
\newcommand{\OPfreezeSYMB}{\cdot}
\newcommand{\OPfreeze}[2]{#1\OPfreezeSYMB#2}
\newcommand{\OPandSYMB}{\wedge}
\newcommand{\OPand}[2]{#1\OPandSYMB#2}
\newcommand{\OPandb}[2]{\begin{array}{l}#1\OPandSYMB\\#2\end{array}}
\newcommand{\OPandbb}[2]{\begin{array}[t]{l}#1\OPandSYMB\\#2\end{array}}
\newcommand{\OPorSYMB}{\vee}
\newcommand{\OPor}[2]{#1\OPorSYMB#2}
\newcommand{\OPimpSYMB}{\rightarrow}
\newcommand{\OPimp}[2]{#1\OPimpSYMB#2}
\newcommand{\OPimpb}[2]{\begin{array}[t]{l}#1\OPimpSYMB\\#2\end{array}}

\newcommand{\OPiff}[2]{#1\leftrightarrow#2}
\newcommand{\OPtaddSYMB}{\bullet}
\newcommand{\OPtadd}[2]{#1\OPtaddSYMB#2}
\newcommand{\OPtzero}{\boldsymbol{0}}
\newcommand{\OPtcompSYMB}{\mathrel{\sim}}
\newcommand{\OPtcomp}[2]{#1\OPtcompSYMB#2}
\newcommand{\OPall}[3]{\forall#1:#2.#3}
\newcommand{\OPallb}[3]{\begin{array}[t]{l}\forall#1:#2.\\#3\end{array}}
\newcommand{\OPex}[3]{\exists#1:#2.#3}

\newcommand{\OPexbb}[3]{\begin{array}[b]{l}\exists#1:#2.\\#3\end{array}}
\newcommand{\OPtltSYMB}{\mathrel{\sqsubset}}
\newcommand{\OPtlt}[2]{#1\OPtltSYMB#2}
\newcommand{\OPtleSYMB}{\sqsubseteq}
\newcommand{\OPtle}[2]{#1\OPtleSYMB#2}
\newcommand{\OPteqSYMB}{\equiv}
\newcommand{\OPteq}[2]{#1\OPteqSYMB#2}
\newcommand{\OPtprevSYMB}{\triangleleft}
\newcommand{\OPtprev}[2]{#1\mathrel{\OPtprevSYMB}#2}
\newcommand{\OPsize}[2]{\mathtt{size}(#1,#2)}
\newcommand{\OPinSYMB}{\in}
\newcommand{\OPin}[2]{#1\OPinSYMB#2}
\newcommand{\OPatSYMB}{@}
\newcommand{\OPat}[2]{\OPatSYMB_{#1}#2}
\newcommand{\ATsend}[3]{\mathtt{send}(#1,#2,#3)}
\newcommand{\ATreceive}[3]{\mathtt{recv}(#1,#2,#3)}
\newcommand{\ATinternal}[2]{\mathtt{internal}(#1,#2)}

\newcommand{\ATcorrect}[1]{\mathtt{correct}(#1)}
\newcommand{\avar}{v_{a}}
\newcommand{\Avar}{v_{A}}
\newcommand{\dvar}{v_{d}}
\newcommand{\tvar}{v_{t}}
\newcommand{\aconst}{c_{a}}

\newcommand{\dconst}{c_{d}}
\newcommand{\interp}[5]{#1,#2,#3,#4\models#5}
\newcommand{\interpp}[4]{#1,#2,#3\models#4}
\newcommand{\interppp}[3]{#1,#2\models#3}
\newcommand{\sem}[2]{\llbracket#1\rrbracket_{#2}}
\newcommand{\tsem}[1]{\llbracket#1\rrbracket}
\newcommand{\csem}[1]{\llbracket#1\rrbracket}
\newcommand{\MAgent}[0]{\mathsf{Agent}}
\newcommand{\MAgents}[0]{\mathsf{Agents}}
\newcommand{\MData}[0]{\mathsf{Data}}
\newcommand{\MTexp}[0]{\mathsf{TimeExp}}
\newcommand{\Mcomparison}[0]{\mathsf{Comparison}}
\newcommand{\da}{a_{d}}
\newcommand{\sa}{a_{s}}
\newcommand{\MDAtom}[0]{\mathsf{DistAtom}}
\newcommand{\MSAtom}[0]{\mathsf{SetAtom}}
\newcommand{\MAtom}[0]{\mathsf{Atom}}

\newcommand{\MForm}[0]{\mathsf{Formula}}
\newcommand{\Mtype}[0]{\mathsf{Type}}
\newcommand{\val}{{\mathcal{V}}}
\newcommand{\run}{r}
\newcommand{\mytuple}[1]{\langle#1\rangle}
\newcommand{\Mworld}[0]{{\cal{W}}}%
\newcommand{\Mglobal}[0]{\mathsf{Global}}

\newcommand{\Matom}[0]{\mathsf{ClosedAtom}}
\newcommand{\Mtvar}[0]{{\mathcal{V}}_{t}}
\newcommand{\Mavar}[0]{{\mathcal{V}}_{a}}
\newcommand{\Maconst}[0]{{\mathcal{C}}_{a}}
\newcommand{\MAvar}[0]{{\mathcal{V}}_{A}}

\newcommand{\Mdvar}[0]{{\mathcal{V}}_{d}}
\newcommand{\Mdconst}[0]{{\mathcal{C}}_{d}}
\newcommand{\Mfun}[2]{#1\rightarrow#2}
\newcommand{\Mfunb}[2]{\begin{array}[t]{l}#1\rightarrow\\#2\end{array}}

\newcommand{\Mforallb}[2]{\begin{array}[t]{l}\forall#1.\\#2\end{array}}
\newcommand{\Mor}[2]{#1\vee#2}
\newcommand{\Mand}[2]{#1\wedge#2}
\newcommand{\Mzero}{\mathtt{0}}
\newcommand{\McompSYMB}{\boldsymbol{\cdot}}
\newcommand{\Mcomp}[2]{#1\McompSYMB#2}
\newcommand{\MleSYMB}{\preceq}
\newcommand{\Mle}[2]{#1\MleSYMB#2}
\newcommand{\MltSYMB}{\prec}
\newcommand{\Mlt}[2]{#1\MltSYMB#2}
\newcommand{\MgeSYMB}{\succeq}
\newcommand{\Mge}[2]{#1\MgeSYMB#2}

\newcommand{\MpredSYMB}{\leftslice}
\newcommand{\Mpred}[2]{#1\MpredSYMB#2}
\newcommand{\MsuccSYMB}{\rightslice}
\newcommand{\Msucc}[2]{#1\MsuccSYMB#2}
\newcommand{\Mlist}[1]{\mathsf{List}(#1)}
\newcommand{\Mset}{\mathsf{Set}}
\newcommand{\DEF}{:\mkern-2mu\equiv}
\newcommand{\OPeventuallySYMB}{\lozenge}
\newcommand{\OPeventually}[1]{\OPeventuallySYMB#1}
\newcommand{\OPeventuallyby}[2]{\OPeventuallySYMB_{#1}#2}
\newcommand{\OPbeforeSYMB}{\blacklozenge}
\newcommand{\OPbefore}[1]{\OPbeforeSYMB#1}
\newcommand{\OPstrictbeforeSYMB}{\blacktriangleleft}
\newcommand{\OPstrictbefore}[1]{{\OPstrictbeforeSYMB}#1}
\newcommand{\OPalwaysSYMB}{\square}
\newcommand{\OPalways}[1]{\OPalwaysSYMB#1}
\newcommand{\OPalwaysby}[2]{\OPalwaysSYMB_{#1}#2}
\newcommand{\OPalwaysbeforeSYMB}{\blacksquare}
\newcommand{\OPalwaysbefore}[1]{\OPalwaysbeforeSYMB#1}
\newcommand{\seq}[3]{#1\vdash#3^{#2}}
\newcommand{\seqb}[2]{#1\vdash#2}
\newcommand{\nseq}[2]{#1\vdash#2}
\newcommand{\hyp}[2]{#1^{#2}}
\newcommand{\hypp}[2]{(#1)^{#2}}
\newcommand{\annot}{\mathsf{a}}
\newcommand{\ANno}{\centerdot}
\newcommand{\ANintcc}[2]{[#1,#2]}
\newcommand{\ANintco}[2]{[#1,#2)}
\newcommand{\ANintoc}[2]{(#1,#2]}
\newcommand{\ANintoo}[2]{(#1,#2)}
\newcommand{\Msub}[3]{#1[#2\backslash#3]}
\newcommand{\calc}{\textsc{tptl-dist}\xspace}

\newcommand{\META}[1]{{\mathit{#1}}}
\newcommand{\comalgo}[1]{\textit{\textcolor{Maroon}{#1}}}

\newcommand{\fanout}{{\cal{F}}}

\newcommand{\rpull}{\mathcal{S}}
\newcommand{\rpush}{\mathcal{P}}
\newcommand{\hb}[2]{\mbox{\texttt{HB}}\left(#1,#2\right)}
\newcommand{\mkmsg}[2]{\langle#1,#2\rangle}
\newcommand{\rrep}{\mathcal{R}}

\newcommand{\tDiffuseSYMB}{\texttt{push}}

\newcommand{\tDiffuse}[3]{\tDiffuseSYMB(#1,#2,#3)}

\newcommand{\storeSYMB}{\texttt{store}}
\newcommand{\store}[1]{\storeSYMB(#1)}

\newcommand{\initializationSYMB}{\texttt{initialisation}}
\newcommand{\initialization}[1]{\initializationSYMB(#1)}

\newcommand{\checkConnectivitySYMB}{\texttt{check-connectivity}}
\newcommand{\checkConnectivity}[1]{\checkConnectivitySYMB(#1)}

\newcommand{\Waxioma}{\textit{$\McompSYMB$-sym}\xspace}
\newcommand{\Waxiomb}{\textit{$\McompSYMB$-assoc}\xspace}
\newcommand{\Waxiomc}{\textit{$\McompSYMB$-unit}\xspace}
\newcommand{\Waxiomd}{\textit{$\Mzero$-initial}\xspace}

\newcommand{\Waxiomf}{\textit{$\MleSYMB$-refl}\xspace}
\newcommand{\Waxiomg}{\textit{$\MleSYMB$-trans}\xspace}
\newcommand{\Waxiomh}{\textit{monotonicity}\xspace}
\newcommand{\Waxiomi}{\textit{sub}\xspace}

\newcommand{\Waxiomk}{\textit{right-density}\xspace}
\newcommand{\Waxioml}{\textit{left-density}\xspace}

\newcommand{\Waxiomo}{\textit{conditional-totality}\xspace}
\newcommand{\WaxiomInd}{\textit{induction}\xspace}

\newcommand{\RULEvarle}{\mbox{\textit{ax}}_{\OPtleSYMB}}
\newcommand{\RULEvar}{\mbox{\textit{ax}}}
\newcommand{\RULElem}{\mbox{\textit{lem}}}
\newcommand{\RULEind}{\mbox{\textit{ind}}}
\newcommand{\RULEthin}{\mbox{\textit{thin}}}

\newcommand{\RULEcut}{\mbox{\textit{cut}}}
\newcommand{\RULEimpR}{{\OPimpSYMB}\mbox{R}}
\newcommand{\RULEimpL}{{\OPimpSYMB}\mbox{L}}
\newcommand{\RULEinR}{{\OPinSYMB}\mbox{R}}
\newcommand{\RULEleTrans}{{\OPtleSYMB}\mbox{\textit{trans}}}
\newcommand{\RULEorL}{{\OPorSYMB}\mbox{L}}
\newcommand{\RULEorRa}{{\OPorSYMB}\mbox{R}_1}
\newcommand{\RULEorRb}{{\OPorSYMB}\mbox{R}_2}
\newcommand{\RULEtrueR}{{\OPtrue}\mbox{R}}
\newcommand{\RULEfalseL}{{\OPfalse}\mbox{L}}
\newcommand{\RULEandL}{{\OPandSYMB}\mbox{L}}
\newcommand{\RULEandR}{{\OPandSYMB}\mbox{R}}

\newcommand{\RULEuntilR}{{\OPuntilSYMB}\mbox{R}}
\newcommand{\RULEuntilL}{{\OPuntilSYMB}\mbox{L}}
\newcommand{\RULEnextR}{{\OPnextSYMB}\mbox{R}}
\newcommand{\RULEnextL}{{\OPnextSYMB}\mbox{L}}
\newcommand{\RULEsinceR}{{\OPsinceSYMB}\mbox{R}}
\newcommand{\RULEsinceL}{{\OPsinceSYMB}\mbox{L}}
\newcommand{\RULEprevR}{{\OPprevSYMB}\mbox{R}}
\newcommand{\RULEprevL}{{\OPprevSYMB}\mbox{L}}
\newcommand{\RULEfreezeR}{{\OPfreezeSYMB}\mbox{R}}
\newcommand{\RULEfreezeL}{{\OPfreezeSYMB}\mbox{L}}
\newcommand{\RULEaddSym}{{\OPtaddSYMB}\mbox{\textit{sym}}}
\newcommand{\RULEiccL}{\mbox{()L}}
\newcommand{\RULEiocL}{\mbox{[)L}}
\newcommand{\RULEicoL}{\mbox{(]L}}
\newcommand{\RULEiooL}{\mbox{[]L}}
\newcommand{\RULEiccR}{\mbox{()R}}
\newcommand{\RULEiocR}{\mbox{[)R}}
\newcommand{\RULEicoR}{\mbox{(]R}}
\newcommand{\RULEiooR}{\mbox{[]R}}
\newcommand{\RULEforallR}{{\forall}\mbox{R}}
\newcommand{\RULEforallL}{{\forall}\mbox{L}}
\newcommand{\RULEexistsR}{{\exists}\mbox{R}}
\newcommand{\RULEexistsL}{{\exists}\mbox{L}}
\newcommand{\RULEalwaysR}{{\OPalwaysSYMB}\mbox{R}}
\newcommand{\RULEalwaysL}{{\OPalwaysSYMB}\mbox{L}}
\newcommand{\RULEeventuallyR}{{\OPeventuallySYMB}\mbox{R}}
\newcommand{\RULEeventuallyL}{{\OPeventuallySYMB}\mbox{L}}
\newcommand{\RULEeventuallybyR}{{\OPeventuallySYMB_{\_}}\mbox{R}}
\newcommand{\RULEeventuallybyL}{{\OPeventuallySYMB_{\_}}\mbox{L}}
\newcommand{\RULEalwaysbyR}{{\OPalwaysSYMB_{\_}}\mbox{R}}
\newcommand{\RULEalwaysbyL}{{\OPalwaysSYMB_{\_}}\mbox{L}}
\newcommand{\RULEstrictbeforeR}{{\OPstrictbeforeSYMB}\mbox{R}}
\newcommand{\RULEstrictbeforeL}{{\OPstrictbeforeSYMB}\mbox{L}}
\newcommand{\RULEbeforeR}{{\OPbeforeSYMB}\mbox{R}}
\newcommand{\RULEbeforeL}{{\OPbeforeSYMB}\mbox{L}}
\newcommand{\RULEalwaysbeforeR}{{\OPalwaysbeforeSYMB}\mbox{R}}
\newcommand{\RULEalwaysbeforeL}{{\OPalwaysbeforeSYMB}\mbox{L}}

\newcommand{\lpar}{(}
\newcommand{\rpar}{)}

\usepackage{booktabs,array}
\ifdefined\DeclareUnicodeCharacter
  \DeclareUnicodeCharacter{22A2}{\ensuremath{\vdash}}
  \DeclareUnicodeCharacter{2192}{\ensuremath{\to}}
  \DeclareUnicodeCharacter{25C7}{\ensuremath{\Diamond}}
  \DeclareUnicodeCharacter{2193}{\ensuremath{\downarrow}}
  \DeclareUnicodeCharacter{2227}{\ensuremath{\land}}
  \DeclareUnicodeCharacter{2228}{\ensuremath{\lor}}
  \DeclareUnicodeCharacter{22C6}{\ensuremath{\star}}
  \DeclareUnicodeCharacter{2080}{\ensuremath{{}_0}}
  \DeclareUnicodeCharacter{2095}{\ensuremath{{}_h}}
\fi
\newcommand{\acode}[1]{\textsf{#1}}               %
\newcommand{\der}{\mathrel{\vdash_{\mathsf{d}}}} %
\newcommand{\fvar}[1]{\mathsf{X}_{#1}}         %
\newcommand{\xvar}[1]{\mathsf{x}_{#1}}            %
\newcommand{\instt}[1]{\llparenthesis #1 \rrparenthesis_{\sigma,\rho}} %
\newcommand{\Just}{\mathsf{just}}
\newcommand{\Nothing}{\mathsf{nothing}}
\newcommand{\Maybe}{\mathsf{Maybe}}

\newcommand{\formalised}{{\color{NavyBlue!75!White}{\raisebox{-0.5pt}{\scalebox{0.8}{\faCog}}}}}
\newcommand{\VGIT}{https://github.com/jaeem006/TPTL-DIST/blob/main/Index2025.lagda}
\newcommand{\VGITa}{https://github.com/jaeem006/TPTL-DIST/blob/main/Index2026.lagda}
\newcommand{\VGITe}{https://github.com/jaeem006/TPTL-DIST/blob/main/Examples.lagda}
\newcommand{\flink}[1]{\href{\VGIT\#L#1}{\formalised}}
\newcommand{\flinka}[1]{\href{\VGITa\#L#1}{\formalised}}

\newcommand{\myurl}[1]{{\footnotesize\textcolor{RoyalBlue}{\url{#1}}}}

\begin{document}

\title{\calc: A Formalised Calculus with Proof Automation for
  Verifying Real-Time Distributed Systems}

\author{Javier Enr\'{i}quez Mendoza\inst{1}%
  \and Sam Speight\inst{2}%
  \and Vincent Rahli\inst{3}%
}
\institute{University of Birmingham \email{jxe275@student.bham.ac.uk}
  \and
  University of Birmingham  \email{s.l.speight@bham.ac.uk}
  \and
  University of Birmingham  \email{V.Rahli@bham.ac.uk}}

\maketitle

\begin{abstract}
We introduce \calc, a temporal logic and labelled sequent calculus for
specifying and verifying real-time distributed systems. \calc is
formalised in Agda to provide machine-checked correctness and tool
support, including automation through a proof search that runs inside
Agda.
The logic extends Timed Propositional Temporal Logic (TPTL) with past
operators, freeze quantification, comparisons between time
expressions, domain-specific atoms, and an induction principle on
time.
We provide a world-based semantics, develop an interval-annotated
calculus, and establish its soundness w.r.t.\ the semantics.
To demonstrate its
expressiveness, we encode standard distributed system patterns, and we
carry out a mechanised proof of a key real-time property of the Pistis
Byzantine broadcast protocol. Our results show that \calc provides a
uniform, mechanised framework for reasoning about temporal properties
of distributed systems.
\keywords{Temporal Logic, Real-Time, Distributed Systems, Byzantine Reliable Broadcast, Agda}
\end{abstract}

\section{Introduction}

Modern systems, such as smart factories, power grids, high-speed
payment processing (e.g., Visa, PayPal), ultra-low-latency 5G
networks, and mission-critical air traffic control,
rely on distributed systems.
In these settings, correctness depends not only on what actions occur,
but when they occur, as delays can lead to failures.
While temporal logics such as LTL and CTL provide well established foundations
for reasoning about system behaviour over time, they lack support
for real-time properties.
Extensions such as TPTL~\cite{henzinger_really_nodate} and Timed
CTL~\cite{Baier+Katoen:2008} incorporate clocks and timing
constraints, and tools
like PRISM~\cite{Prism:2011} and
TLA+~\cite{lamport_specifying_2003,Lamport:real:2005,TLA:2012} support modelling
and model checking of real-time and distributed
systems~\cite{Norman+Parker+Zou:formats:2015,Oxford+Parker+Ryan:cns:2020,%
Lu+Merz+Weidenbach:forte:2011,Joshi+al:fmsd:2003,Bolosky+Douceur+Howell:sigops:2007,%
Newcombe:asm:2014,Newcombe+al:cacm:2015,Chand+Liu+Stoller:fm:2016,Gao+al:fm:2021}.
However, existing formalisms differ from our approach in important
ways and face key limitations.
(1) Our work leverages a full Agda formalisation, enabling
machine-checked reasoning and providing a foundation for verified tool
support.
(2) Logics like TPTL and Timed CTL cannot express common distributed
system events such as message send/receive or agent correctness, and
they lack inductive reasoning over time.
(3) Tools like PRISM and TLA+ require exhaustively exploring all
possible states of the system, which is not always feasible.
(4) Existing approaches separate clock reasoning from the
communication structure of distributed systems.

To complement the current landscape of approaches, we introduce \calc,
a fully formalised temporal logic and labelled sequent calculus
designed specifically for reasoning about real-time fault-tolerant
distributed protocols. The logic extends TPTL with past operators,
freeze quantification, a general comparison operator on time
expressions, and atomic predicates encoding message transmission,
internal events, and agent correctness.

\calc's rules are annotated with time intervals and the semantics is
given over an abstract world structure equipped with an induction
principle for temporal reasoning.
A central goal of \calc is to provide a uniform framework to state and
reason about both temporal and distributed aspects of a protocol.
All definitions, rules, and meta-theoretical results are formalised in
Agda:~\myurl{\VGIT} and \myurl{\VGITa}.
We use the symbol~\formalised\xspace to link to the formalization.

\paragraph{Contributions}
Our contributions are as follows, demonstrating that \calc provides a
principled and mechanised foundation for reasoning about real-time
distributed systems:
\begin{itemize}
\item \calc, a new logic for real-time distributed systems, extending TPTL with
  past operators, quantifiers, comparisons on time expressions, and
  fault-tolerant distributed system operators.
\item A sequent calculus for \calc that supports reasoning with intervals,
  which is sound for the intended semantics.
\item A general semantic framework based on ordered worlds, enabling a principled
  induction rule for unbounded temporal reasoning.
\item A mechanization in Agda, covering the logic, calculus and meta-theory.
\item A modular design of the Pistis real-time reliable broadcast
  protocol~\cite{Kozhaya+Decouchant+Rahli+Verissimo:tpds:2021}.
\item A mechanised proof of Pistis' timeliness property, showcasing
  \calc's support for real-time constraints, message-passing patterns,
  and induction.
\item We automate proofs in \calc{} with a proof search that runs
  inside Agda and whose results are checked by Agda’s type checker.
\end{itemize}
While the calculus and its application to the Pistis protocol were
presented in the conference
paper~\cite{Mendoza+Speight+Rahli:tptl-dist:forte:2026}, the
automation presented in Sec.~\ref{sec:automation} is new.

\section{Related Work}

Several proof systems have been proposed for LTL-like logics. However,
the until operator is known to pose difficulties due to its dual
character of being both universal and existential.
A labelled natural deduction system was proposed in~\cite{Volpe:2010}
to treat those difficulties. Simplifying and generalising that system
(see Sec.~\ref{sec:tptl-dist}), we propose a labelled sequent calculus
that allows reasoning about TPTL's temporal operators, in particular
until and freeze, in a uniform way.

\paragraph{TPTL}
The syntax of TPTL is as follows, where $\OPuntilSYMB$ and
$\OPnextSYMB$ are the usual ``until'' and ``next'' LTL operators;
$p$ is an atomic proposition;
$x$ is a clock variable;
$c$ is a clock expression;
$\sim~\in\{\leq,<,=,>,\geq\}$ allows comparing a clock variable with a
clock expression;
and the \emph{freeze} formula $\OPfreeze{x}{\varphi}$ binds the current
time to~$x$ in~$\varphi$:
\[
\begin{small}
    \begin{array}{lrl}
        \varphi&::=& p\mid
            \varphi\land\varphi\mid
            \OPneg{\varphi}\mid
            \OPuntil{\varphi}{\varphi}\mid
            \OPnext{\varphi}\mid
            \OPop{x}{c}\mid
            \OPfreeze{x}{\varphi}
    \end{array}
\end{small}
\]

The \emph{eventually} and \emph{always} operators are defined as
usual: $\OPeventuallySYMB{\varphi}\DEF\OPuntil{\OPtrue}{\varphi}$ and
$\OPalways{\varphi}\DEF\OPneg{\OPeventually{\OPneg{\varphi}}}$, while
the freeze operator allows capturing that~$\varphi$ eventually
holds by time~$c$:
$\OPfreeze{x}{\OPeventually{(\OPand{x\le{c}}{\varphi})}}$.
To understand this, consider TPTL's semantics over the discrete
notion of time given by $\mathbb{N}$. A formula's semantics is
given in terms of a mapping~$\rho$ from~$\mathbb{N}$ to states, a
point in time $i\in\mathbb{N}$, and a variable valuation~$v$, where a
state~$\rho_n$ is a pair
of a valuation on
atoms~$\sigma_n$ and a timestamp~$\tau_n$:
\[
\begin{small}
\begin{array}{c}
    \begin{array}{lrl}
        \rho,i,v\models p&\iff& p\in\sigma_i\\
        \rho,i,v\models\neg\varphi&\iff&\rho,i,v\not\models\varphi\\
        \rho,i,v\models x\sim c&\iff&\tau_i-v(x)\sim c\\
    \end{array}
    \hspace{0.2in}
    \begin{array}{lrl}
        \rho,i,v\models\varphi\land\psi&\iff&\rho,i,v\models \varphi\mbox{ and }\rho,i,v\models\psi\\
        \rho,i,v\models\bigcirc\varphi&\iff&\rho,i+1,v\models\varphi\\
        \rho,i,v\models x\cdot\varphi&\iff&\rho,i,v[x\mapsto\tau_i]\models\varphi\\
    \end{array}
    \\
    \begin{array}{lrl}
        \rho,i,v\models\OPuntil{\varphi}{\psi}&\iff
        &\exists{j\geq{i}}. (\rho,j,v\models\psi)\wedge\forall{k\in[i,j)}.(\rho,k,v\models\varphi)\\
    \end{array}
\end{array}
\end{small}
\]
The semantics of ``$\varphi$ eventually holds by time~$c$'', i.e.,
$x\cdot{\lozenge(x\le{c}\wedge\varphi)}$ w.r.t.\ $\rho,i,v$ states
that there exists a time $\tau_j\leq{\tau_i+c}$ at which $\varphi$ is
true, for some $j\geq{i}$.
This is achieved by ``freezing'' the current time $\tau_i$, by mapping
$x$ to $\tau_i$ in the variable valuation $v$; by jumping to a later
time $\tau_j$ thanks to the $\lozenge$ operator; and finally the
constraint $x\le{c}$ guarantees that $\tau_j\le{\tau_i+c}$.

Despite its expressive power, TPTL does not allow formalizing
properties of distributed systems such as those relevant to the Pistis
real-time Byzantine-resilient reliable broadcast
protocol~\cite{Kozhaya+Decouchant+Rahli+Verissimo:tpds:2021}, lacking
support for reasoning about agents, messages, communication, etc.
\calc is designed to address these points, as well as TPTL's lack of
support for important time-related reasoning mechanisms, such as
reasoning about past properties in addition to future properties, and
reasoning inductively over time.

Note that \calc aims to support the specification and
verification of systems that are required to perform critical actions
in a timely fashion.
As such, it allows modelling a \emph{real} or \emph{physical}, quantitative
notion of time, in addition to a \emph{logical}, qualitative notion of
time~\cite{Kopetz:1997,Kopetz:2014}.%

Other approaches with similar motivation, i.e., formally verifying
timing properties of systems, but using different techniques have been
proposed in the literature.
We illustrate
below how our approach relates and differs.

\paragraph{TLA+: Temporal Logic of Actions}
The primary distinction with TLA+ lies in its treatment of time. Our
system integrates temporal constructs directly into the language's
syntax through its freeze and comparison operators. This
representation of time allows for simple and intuitive modelling of
temporal conditions.
By contrast, TLA+ relies on temporal operators such as always and
eventually to express time-dependent properties. Although it is
feasible to capture a quantitative notion of time in TLA+ using
variables that represent time, these variables are treated
as ordinary state variables and are not inherently connected to the
temporal operators.
As a result, while providing great flexibility, the burden of
correctly managing time falls on the user explicitly updating
those variables, as opposed to the automated management of time
in \calc. %

\paragraph{PRISM: Probabilistic Model Checking}
PRISM is primarily designed for the verification
of probabilistic properties. It is based on probabilistic temporal logics
such as PCTL (Probabilistic Computation Tree Logic)~\cite{HJ94,BdA95}
and CSL (Continuous Stochastic Logic)~\cite{ASSB96,BKH99}. In addition
to including temporal mechanisms that enable reasoning about
time-related properties, PRISM's core strength lies in its ability to
handle systems characterized by randomness, uncertainty, and
nondeterminism.
\calc complements PRISM's approach, whose model checking
engine imposes some restrictions on the real-time systems and
properties that can be verified, providing flexibility to the users
thanks to its calculus.

\paragraph{Summary}
TLA+ lets users manage time flexibly through variables, requiring
careful handling, while PRISM takes a more restricted approach to
enable effective model checking.
\calc provides
primitives that do not require careful management of variables,
while still providing some flexibility when stating time-related
properties, along with a calculus to prove those properties.

\section{\calc}
\label{sec:tptl-dist}

We now present \calc, an extension of TPTL +
Past~\cite{bozzelli_alternating-time_2020}, with quantifiers, a more
general comparison operator, a ``discrete'' semantics, and a sequent
calculus.

\subsection{Syntax}

To reason about distributed systems, which are composed of nodes, or
agents, communicating by exchanging messages, we fix some types ($T$)
to allow quantifying over agents ($\LtyAgent$), sets of agents
($\LtyAgents$), and data ($\LtyData$), where $\LtyData$ is used to
capture the information exchanged between agents~\flink{40}:
$$T \in \Mtype ::= \LtyAgent\mid\LtyAgents\mid\LtyData$$

Let~$\avar$ range over a set~$\Mavar$ of variables of type~$\LtyAgent$
and~$\aconst$ range over a set~$\Maconst$ of constants of
type~$\LtyAgent$;
let~$\Avar$ range over a set~$\MAvar$ of variables of type~$\LtyAgents$; and
let~$\dvar$ range over a set~$\Mdvar$ of variables of type~$\LtyData$
and~$\dconst$ range over a set~$\Mdconst$ of constants of
type~$\LtyData$.
Expressions of types $\LtyAgent$~\flink{43}, $\LtyAgents$~\flink{46}, and
$\LtyData$~\flink{49} are defined as follows:
$$\begin{array}{l@{\hspace{0.15in}}l@{\hspace{0.15in}}l}
\begin{array}{lrl}
  i,j \in \MAgent & ::= & \avar\mid\aconst\\
\end{array}
&
\begin{array}{lrl}
  A \in \MAgents & ::= & \Avar\mid[i_1,\dots,i_n]\\
\end{array}
&
\begin{array}{lrl}
  d \in \MData & ::= & \dvar\mid\dconst\\
\end{array}
\end{array}$$

\calc also includes expressions to capture time constraints,
 where~$\tvar$ ranges over a set~$\Mtvar$ of time variables,
$\OPtzero$ is a constant, capturing an empty amount of time, and
$\OPtaddSYMB$ is a binary operation, intuitively adding two time
expressions~\flink{52}:
$$t \in \MTexp ::= \tvar\mid\OPtzero\mid\OPtadd{t}{t}$$

To compare time expressions, we further introduce the
following operators~\flink{55}:
$$\OPtcompSYMB~ \in \Mcomparison ::= ~\OPtltSYMB~\mid~\OPteqSYMB~\mid\OPtprevSYMB$$
Intuitively, $\OPtlt{t_1}{t_2}$ states
that~$t_1$ is strictly less than~$t_2$, $\OPteq{t_1}{t_2}$ states
that~$t_1$ and~$t_2$ are equal, and $\OPtprev{t_1}{t_2}$ states
that~$t_1$ is $t_2$'s immediate predecessor. %

To reason about distributed systems, one must reason about
the knowledge acquired by agents over time.
The two main approaches are the \emph{implicit} or \emph{external}
representation of knowledge, which is externally assigned to
agents; and the \textit{explicit} representation, where knowledge is explicitly
``stored'' in states, allowing agents to compute and act on their
knowledge~\cite{fagin_reasoning_1995}.%
We take an explicit approach, introducing a number of atomic
expressions that capture knowledge through the exchange of
messages between agents.
A \emph{distributed} atom~$a$ is either
a \emph{send} atom $\ATsend{i}{d}{A}$, stating that agent~$i$ sent the
data~$d$ to the set of agents~$A$; or a \emph{receive} atom
$\ATreceive{i}{d}{j}$, stating that agent~$i$ received the data~$d$
from agent~$j$; or an \emph{internal event} atom $\ATinternal{i}{d}$,
stating that~$d$ happened at agent~$i$; or a \emph{correct} atom
$\ATcorrect{i}$, stating that agent~$i$ is correct~\flink{58}:
$$\begin{array}{lrl}
  \da \in \MDAtom & ::= & %
  \ATsend{i}{d}{A}\mid\ATreceive{i}{d}{j}
   \mid\ATinternal{i}{d}\mid\ATcorrect{i}
\end{array}$$

Furthermore, \calc includes atoms to reason about sets of agents
as follows, where $n\in\mathbb{N}$, $\OPsize{A}{n}$
states that the collection of agents $A$ is of size $n$, and
$\OPin{i}{A}$ states that agent~$i$ is in the collection of
agents~$A$~\flink{61}:
$$\begin{array}{lrl}
  \sa \in \MSAtom & ::= &
            \OPsize{A}{n}\mid
            \OPin{i}{A}
\end{array}$$

An atom is then either a \emph{distributed} or a \emph{set} atom:
$a\in\MAtom::=\da\mid\sa$.

\calc's syntax includes: (1)~propositional connectives; (2)~past and
future LTL operators; (3)~a freeze operator along with a comparison
operator on time expressions; (4)~quantifiers; (5)~atoms to reason
about distributed systems.
It is defined as follows, where~$\OPsinceSYMB$ and~$\OPprevSYMB$ are
the ``since'' and ``yesterday'' past counterparts of $\OPuntilSYMB$
and $\OPnextSYMB$,
and $v$ is a variable of type $T$ in $\OPall{v}{T}{\varphi}$ and
$\OPex{v}{T}{\varphi}$~\flink{64}:
\[
\begin{array}{lrl}
        \varphi \in \MForm
         &::=& a\mid
            \OPtrue\mid
            \OPfalse\mid
            \OPand{\varphi}{\varphi}\mid
            \OPor{\varphi}{\varphi}\mid
            \OPimp{\varphi}{\varphi}\mid
            \OPuntil{\varphi}{\varphi}\mid
            \OPnext{\varphi}\\
            & \mid &
            \OPsince{\varphi}{\varphi}\mid
            \OPprev{\varphi}\mid
            \OPtcomp{t}{t}\mid
            \OPfreeze{\tvar}{\varphi}\mid
            \OPall{v}{T}{\varphi}\mid
            \OPex{v}{T}{\varphi}\\
\end{array}
\]

We write $\OPall{v_1,\dots,v_n}{T}{\varphi}$ for
$\OPall{v_1}{T}{\dots\OPall{v_n}{T}{\varphi}}$, and similarly for
$\exists$.
Negation is defined as usual: $\OPneg{\varphi}\DEF\OPimp{\varphi}{\OPfalse}$.
We recall the definitions of \textit{eventually} and \textit{always} (in white)
as well as their past counterparts (in black)~\flink{67}:
\[
\begin{array}{l@{\hspace{0.2in}}l@{\hspace{0.2in}}l@{\hspace{0.2in}}l}
  \OPeventually{\varphi} \DEF \OPuntil{\OPtrue}{\varphi}
  &
  \OPalways{\varphi} \DEF \OPneg{\OPeventually{\OPneg{\varphi}}}
  &
  \OPbefore{\varphi} \DEF \OPsince{\OPtrue}{\varphi}
  &
  \OPalwaysbefore{\varphi} \DEF \OPneg{\OPbefore{\OPneg{\varphi}}}
\end{array}
\]

Thanks to \calc's general comparison operator $\OPtcomp{t_1}{t_2}$, a
time-bounded version of $\OPeventuallySYMB$ can be defined more
straightforwardly than in TPTL:
$\OPeventuallyby{t}{\varphi}\DEF\OPfreeze{\tvar}{\OPeventually{(\OPfreeze{\tvar'}{\OPand{\OPtle{\tvar'}{(\OPtadd{\tvar}{t})}}{\varphi}})}}$,
stating that~$\varphi$ happens sometime in the future by time~$t$,
where $\OPtle{t_1}{t_2}$ stands for
$\OPor{\OPtlt{t_1}{t_2}}{\OPteq{t_1}{t_2}}$.
Similarly, we define a time-bounded version of the $\OPalwaysSYMB$
operator:
$\OPalwaysby{t}{\varphi}\DEF\OPfreeze{\tvar}{\OPalways{(\OPfreeze{\tvar'}{\OPimp{\OPtle{\tvar'}{(\OPtadd{\tvar}{t})}}{\varphi}})}}$
stating that~$\varphi$ always holds between now and~$t$.
Furthermore, we define the following \emph{strict} version of
the \emph{sometimes before} operator $\OPbeforeSYMB$ as follows:
$\OPstrictbefore{\varphi}\DEF\OPprev{(\OPbefore{\varphi})}$, stating
that $\varphi$ happened sometime in the past, but not now~\flink{75}.

\subsection{Forwarding Example}
\label{sec:application-simple}

We now illustrate aspects of the language using a simple forwarding
example, demonstrating how to capture message passing patterns in
distributed systems, where agents exchange information by
sending/receiving messages (see Appx.~\ref{sec:further-examples} for
further examples).
In this example, messages are relayed between agents of a synchronous
system.
Sec.~\ref{sec:examples-of-simple-proofs} then proves a property of
this system in detail, illustrating the use of \calc's rules presented
in Sec.~\ref{sec:sequent-calculus}.

\begin{figure}[!t]
\begin{center}
  \includegraphics[width=0.5\textwidth]{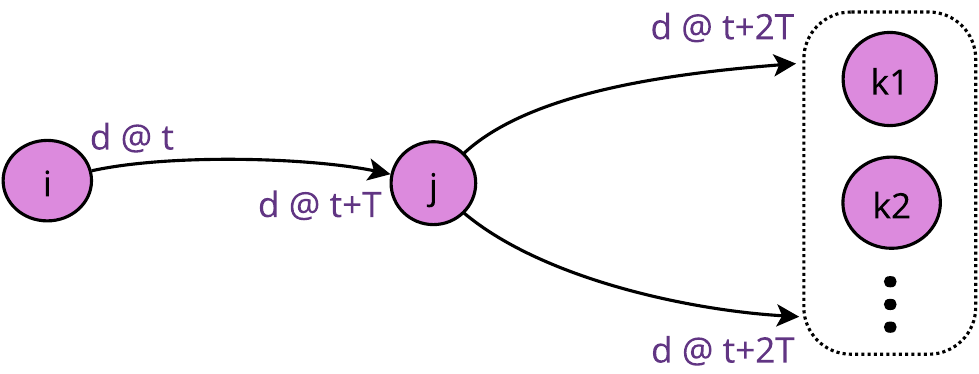}
\vspace*{-0.2in}
\end{center}
\caption{Forwarding example}
\label{fig:forwarding example}
\end{figure}

As illustrated in Fig.~\ref{fig:forwarding example}, in this example,
an agent~$i$ sends a message~$d$ at time~$t$ to another agent~$j$, which
then forwards~$d$ to all other agents. We are then guaranteed that any
agent would have received that message by time $t+2T$ assuming that the
maximum transmission delay between agents is~$T$.

All agents are assumed to be correct and communication is assumed to
be synchronous, ensuring that all sent messages are received within a
time interval~$T$, captured by the following formula stating that if
an agent sends a message~$d$ at time~$t$ to a collection of agents,
those agents will receive~$d$ by time $t+T$~\flink{138}:
\begin{equation}
\begin{small}
\OPall{i,j}{\LtyAgent}{
  \OPall{d}{\LtyData}{
    \OPallb{A}{\LtyAgents}{\hspace*{-0.8in}
        \OPalways{(\OPimp{\ATsend{i}{d}{A}}{\OPeventuallyby{T}{(\OPimp{\OPin{j}{A}}{\ATreceive{j}{d}{i}})}})}
}}}
\end{small}
\label{equ:simple-example-hyp1}
\end{equation}
Additionally, we assume that agents immediately forward the
information they receive to all other agents, captured by the
following formula stating that whenever~$j$ receives a message it
immediately sends it to $k$, and this for any~$k$~\flink{141}:
\begin{equation}
\begin{small}
\OPall{i,j,k}{\LtyAgent}{
  \OPall{d}{\LtyData}{
        \OPalways{(\OPimp{\ATreceive{j}{d}{i}}{\ATsend{j}{d}{[k]}})}
}}
\end{small}
\label{equ:simple-example-hyp2}
\end{equation}
As shown in Sec.~\ref{sec:examples-of-simple-proofs}, using \calc's
calculus presented below, one can then derive the following formula
from the above assumptions, stating that if an agent~$i$ sends a
message~$d$ to some agent~$j$, then any agent~$k$ will also receive
$d$ within $2T$ (where $2T$ is simply $\OPtadd{T}{T}$)~\flink{144}:
\begin{equation}
\begin{small}
\OPall{i,j,k}{\LtyAgent}{
  \OPall{d}{\LtyData}{
        \OPalways{(\OPimp{\ATsend{i}{d}{[j]}}{\OPeventuallyby{2T}{\ATreceive{k}{d}{j}}})}
}}
\end{small}
\label{equ:simple-example-concl}
\end{equation}

\subsection{Semantics}
\label{sec:semantics}

\begin{figure*}[!t]
\begin{small}
$$\begin{array}{c}
\begin{array}{l@{\hspace{0.1in}}l}
\mbox{\Waxioma} & \forall{w_1,w_2:\Mworld}.\Mcomp{w_1}{w_2}=\Mcomp{w_2}{w_1}\\
\mbox{\Waxiomb} & \forall{w_1,w_2,w_3:\Mworld}.\Mcomp{w_1}{(\Mcomp{w_2}{w_3})}=\Mcomp{(\Mcomp{w_1}{w_2})}{w_3}\\
\mbox{\Waxiomg} & \forall{w_1,w_2,w_3:\Mworld}.\Mfun{\Mle{w_1}{w_2}}{\Mfun{\Mle{w_2}{w_3}}{\Mle{w_1}{w_3}}}\\
\mbox{\Waxiomc} & \forall{w:\Mworld}.\Mcomp{\Mzero}{w}=w\\
\mbox{\Waxiomf} & \forall{w:\Mworld}.\Mle{w}{w}\\
\mbox{\Waxiomd} & \forall{w:\Mworld}.\Mle{\Mzero}{w}\\
\mbox{\Waxiomh} & \forall{w_1,w_2,w_3,w_4:\Mworld}.\Mfun{\Mle{w_1}{w_2}}{\Mfun{\Mle{w_3}{w_4}}{\Mle{\Mcomp{w_1}{w_3}}{\Mcomp{w_2}{w_4}}}}\\
\mbox{\Waxiomi} & \forall{w_1,w_2:\Mworld}.\Mfun{\Mpred{w_1}{w_2}}{\Mle{w_1}{w_2}}\\
\mbox{\Waxiomk} & \forall{w_1,w_2:\Mworld}.\Mfun{\Mle{w_1}{w_2}}{\Mor{w_1=w_2}{\exists{w:\Mworld}.\Mand{\Mle{w_1}{w}}{\Mpred{w}{w_2}}}}\\
\mbox{\Waxioml} & \forall{w_1,w_2:\Mworld}.\Mfun{\Mle{w_1}{w_2}}{\Mor{w_1=w_2}{\exists{w:\Mworld}.\Mand{\Mpred{w_1}{w}}{\Mle{w}{w_2}}}}\\
\mbox{\Waxiomo} & \forall{w_1,w_2,w:\Mworld}.\Mfun{\Mle{w_1}{w}}{\Mfun{\Mle{w_2}{w}}{\Mor{\Mle{w_1}{w_2}}{\Mle{w_2}{w_1}}}}\\
\WaxiomInd &
\Mforallb
{P:\Mfun{\Mworld}{\Mset}}
{\Mfunb
  {(\forall{w:\Mworld}.{\Mfun{(\forall{u,v:\Mworld}.{\Mfun{\Mle{u}{v}}{\Mfun{\Mpred{v}{w}}{P(u)}}})}{P(w)}})}
  {\forall{w:\Mworld}.{P(w)}}
}
\end{array}
\end{array}$$
\end{small}
\vspace{-0.1in}
\caption{World axioms}
\label{fig:world-axioms}
\end{figure*}

We now adapt and extend TPTL's semantics to our calculus, in
particular to account for domain-specific atoms and a more general
comparison operator between time expressions.
Formulas are interpreted w.r.t.\ a collection of worlds, which
provides meaning to time expressions.
The set of worlds $\Mworld$ includes a constant $\Mzero$ and a binary
operation~$\McompSYMB$ (the semantic counterparts of the time
expressions in $\MTexp$) and is equipped with two relations $\MleSYMB$
and $\MpredSYMB$ (the semantic counterparts of the relations in
$\Mcomparison$) satisfying the axioms in
Fig.~\ref{fig:world-axioms}~\flink{91}.

The axioms \Waxioma, \Waxiomb, \Waxiomc, \Waxiomf, \Waxiomg, \Waxiomh
and \Waxiomd make $\Mworld$ into a positively ordered (commutative)
monoid~\cite{WEHRUNG199243} (though note that we do not require
antisymmetry). Ordered commutative monoids have been exhibited as the
mathematical structures of resource combination and
convertibility~\cite{FRITZ_2017}.
Later, we will interpret time expressions as worlds, and, as such,
may think of time as a resource.

We think of $\MpredSYMB$ as capturing immediate predecessor in the
$\MleSYMB$ relation; hence the \Waxiomi axiom simply expresses that
the former is a subrelation of the latter. Axiom \Waxiomk is used, in
particular, along with the induction axiom, to derive the induction
rule presented in Sec.~\ref{sec:sequent-calculus}, which is key to
proving our main application in
Lem.~\ref{lem:unbounded-delivery-imples-bounded-delivery}.
Axiom \Waxiomo, in the presence of \Waxiomf
and \Waxiomh, is equivalent to totality, by setting $w=\Mcomp{w_1}{w_2}$.

The \Waxiomk and \Waxioml axioms are in particular used to prove the
left-to-right and right-to-left implications of
$\OPiff{(\OPbefore{(\OPprev{\varphi})})}{(\OPprev{(\OPbefore{\varphi})})}$,
respectively. Interestingly, only the former was necessary in
Sec.~\ref{sec:application}.

\begin{figure*}[!t]
\begin{small}
\[
\begin{array}{l}
  \begin{array}{lrl}
    \multicolumn{3}{l}{\fbox{$\tsem{\_} : \Mtype \rightarrow \Mset$}}\\
    \tsem{\LtyAgent}&\DEF&\Maconst\\
    \tsem{\LtyAgents}&\DEF&\Mlist{\Maconst}\\
    \tsem{\LtyData}&\DEF&\Mdconst\\
  \end{array}
  \hspace{0.1in}
  \begin{array}{lrl}
    \multicolumn{3}{l}{\fbox{$\sem{\_}{\val} : \MTexp \rightarrow \Mworld$}}\\
    \sem{\tvar}{\val}&\DEF&\val(\tvar)\\
    \sem{\OPtzero}{\val}&\DEF&\Mzero\\
    \sem{\OPtadd{t_1}{t_2}}{\val}&\DEF&\Mcomp{\sem{t_1}{\val}}{\sem{t_2}{\val}}\\
  \end{array}
  \hspace{0.1in}
  \begin{array}{lrl}
    \multicolumn{3}{l}{\fbox{$\csem{\_} : \Mcomparison \rightarrow \Mfun{\Mworld}{\Mfun{\Mworld}{\Mset}}$}}\\
    \csem{\OPtltSYMB}   & \DEF & \mathrel{\MltSYMB} \\
    \csem{\OPteqSYMB}   & \DEF & \mathrel{=} \\
    \csem{\OPtprevSYMB} & \DEF & \mathrel{\MpredSYMB} \\
  \end{array}
  \\ \\
  \begin{array}{lrl}
    \multicolumn{3}{l}{\fbox{$\sem{\_}{\val} : \MAgents \rightarrow \Mlist{\Maconst}$}}\\
    \sem{\Avar}{\val}&\DEF&\val(\Avar)\\
    \sem{[i_1,\dots,i_n]}{\val}&\DEF&[\sem{i_1}{\val},\dots,\sem{i_n}{\val}]\\
  \end{array}
  \hspace{0.1in}
  \begin{array}{lrl}
    \multicolumn{3}{l}{\fbox{$\sem{\_}{\val} : \MAgent \rightarrow \Maconst$}}\\
    \sem{\avar}{\val}&\DEF&\val(\avar)\\
    \sem{\aconst}{\val}&\DEF&\aconst\\
  \end{array}
  \hspace{0.1in}
  \begin{array}{lrl}
    \multicolumn{3}{l}{\fbox{$\sem{\_}{\val} : \MData \rightarrow \Mdconst$}}\\
    \sem{\dvar}{\val}&\DEF&\val(\dvar)\\
    \sem{\dconst}{\val}&\DEF&\dconst\\
  \end{array}
  \\ \\
  \begin{array}{lrl}
    \multicolumn{3}{l}{\fbox{$\sem{\_}{\val} : \MDAtom \rightarrow \Matom$}}\\
    \sem{\ATsend{i}{d}{A}}{\val} & \DEF & \ATsend{\sem{i}{\val}}{\sem{d}{\val}}{\sem{A}{\val}}\\
    \sem{\ATreceive{i}{d}{j}}{\val} & \DEF & \ATreceive{\sem{i}{\val}}{\sem{d}{\val}}{\sem{j}{\val}}\\
    \sem{\ATinternal{i}{d}}{\val} & \DEF & \ATinternal{\sem{i}{\val}}{\sem{d}{\val}}\\
    \sem{\ATcorrect{i}}{\val} & \DEF & \ATcorrect{\sem{i}{\val}}\\
  \end{array}
\end{array}
\]
\end{small}
\vspace{-0.15in}
\caption{Semantics of types, time expressions, comparison operators,
  agents, collections of agents, data, and distributed atoms}
\label{fig:semantics-types}
\end{figure*}

Following~\cite{fagin_reasoning_1995,herzig_epistemic_2019,baltag_logic_2014,baltag_logic_nodate},
formulas are interpreted w.r.t.\ a model, which is a tuple of the form
$\mytuple{\run,\pi,w,\val}$ consisting of (1)~a run~$\run$, (2)~an
interpretation function~$\pi$, which assigns meaning to atomic
formulas, (3)~a world~$w\in\Mworld$, and (4)~a variable
valuation~$\val$ assigning meaning to free variables.
A run~$\run$ is a function in $\Mfun{\Mworld}{\Mglobal}$, where
$\Mglobal$ is a global state, i.e., an element of $\Mlist{\Matom}$, where
$\Matom$ is the type of closed atoms (closed elements of $\MAtom$).
An interpretation is a function in
$\Mfun{\Mglobal}{\Mfun{\Matom}{\Mset}}$, which checks whether a given
atom is true in a given global state.
A variable valuation is a function
in $(\Mavar\rightarrow\Maconst)\cup(\MAvar\rightarrow\Mlist{\Maconst})\cup(\Mdvar\rightarrow\Mdconst)\cup(\Mtvar\rightarrow\Mworld)$.
We write $\val(X_1,\dots,X_n)$ to indicate that $\val$ is a variable
valuation with domain including the free variables of the syntactic
expressions $X_1$ to $X_n$.
We also write $\Mlt{w_1}{w_2}$ for
$\exists{w:\Mworld}.{\Mand{\Mpred{w_1}{w}}{\Mle{w}{w_2}}}$.

Fig.~\ref{fig:semantics-types}~\flink{94} introduces the semantics of time
expressions $\sem{t}{\val}\in\Mworld$, types $\tsem{T}\in\Mset$,
atoms $\sem{a}{\val}\in\Matom$, agents $\sem{i}{\val}\in\Maconst$,
collections of agents $\sem{A}{\val}\in\Mlist{\Maconst}$, data
$\sem{d}{\val}\in\Mdconst$, and comparison operators
$\csem{\OPtcompSYMB}\in\Mfun{\Mworld}{\Mfun{\Mworld}{\Mset}}$,
while
Fig.~\ref{fig:semantics}~\flink{127} defines the semantics of
formulas.
In particular, the semantics of an atom $\da$ checks that the
interpretation $\sem{\da}{\val}$ is in the global state returned
by~$r$ at world~$w$.

\begin{figure*}[!t]
\begin{small}
\[
\begin{array}{c}
  \begin{array}{lrl}
        \interp{\run}{\pi}{w}{\val}{\da} & \iff & \pi(r(w))(\sem{\da}{\val})\\
        \interp{\run}{\pi}{w}{\val}{\OPin{i}{A}}&\iff&\sem{i}{\val}\in\sem{A}{\val}\\
        \interp{\run}{\pi}{w}{\val}{\OPsize{A}{n}}&\iff&|\sem{A}{\val}|=n\\
        \interp{\run}{\pi}{w}{\val}{\OPtrue} & \iff & \mbox{True}\\
        \interp{\run}{\pi}{w}{\val}{\OPfalse} & \iff & \mbox{False}\\
        \interp{\run}{\pi}{w}{\val}{\OPand{\varphi}{\psi}} & \iff & \interp{\run}{\pi}{w}{\val}{\varphi}\mbox{ and }\interp{\run}{\pi}{w}{\val}{\psi}\\
        \interp{\run}{\pi}{w}{\val}{\OPor{\varphi}{\psi}} & \iff & \interp{\run}{\pi}{w}{\val}{\varphi}\mbox{ or }\interp{\run}{\pi}{w}{\val}{\psi}\\
        \interp{\run}{\pi}{w}{\val}{\OPimp{\varphi}{\psi}} & \iff & \interp{\run}{\pi}{w}{\val}{\varphi}\mbox{ implies }\interp{\run}{\pi}{w}{\val}{\psi}\\
        \interp{\run}{\pi}{w}{\val}{\OPall{v}{T}{\varphi}}&\iff&\interp{\run}{\pi}{w}{\val[v\mapsto{z}]}{\varphi}\mbox{ for all }z\in \tsem{T}\\
        \interp{\run}{\pi}{w}{\val}{\OPex{v}{T}{\varphi}}&\iff&\interp{\run}{\pi}{w}{\val[v\mapsto{z}]}{\varphi}\mbox{ for some }z\in \tsem{T}\\
        \interp{\run}{\pi}{w}{\val}{\OPop{t_1}{t_2}}&\iff&\sem{t_1}{\val}\mathrel{\csem{\sim}}\sem{t_2}{\val}\\
        \interp{\run}{\pi}{w}{\val}{\OPfreeze{\tvar}{\varphi}}&\iff&\interp{\run}{\pi}{w}{\val[\tvar\mapsto{w}]}{\varphi}\\
        \interp{\run}{\pi}{w}{\val}{\OPuntil{\varphi}{\psi}}&\iff
        &\begin{array}[t]{l}\exists{\Mge{w'}{w}}.(\interp{\run}{\pi}{w'}{\val}{\psi})\wedge\\\hspace{0.2in}\forall{w''}.\Mle{w}{\Mlt{w''}{w'}}\rightarrow(\interp{\run}{\pi}{w''}{\val}{\varphi})\end{array}\\
        \interp{\run}{\pi}{w}{\val}{\OPnext{\varphi}}&\iff&\exists{\Msucc{w'}{w}}.\interp{\run}{\pi}{w'}{\val}{\varphi}\\
        \interp{\run}{\pi}{w}{\val}{\OPsince{\varphi}{\psi}}&\iff
        &\begin{array}[t]{l}\exists{\Mle{w'}{w}}.(\interp{\run}{\pi}{w'}{\val}{\psi})\wedge\\\hspace{0.2in}\forall{w''}.\Mlt{w'}{\Mle{w''}{w}}\rightarrow(\interp{\run}{\pi}{w''}{\val}{\varphi})\end{array}\\
        \interp{\run}{\pi}{w}{\val}{\OPprev{\varphi}}&\iff&\exists{\Mpred{w'}{w}}.\interp{\run}{\pi}{w'}{\val}{\varphi}\\
    \end{array}
\end{array}
\]
\end{small}
\vspace{-0.15in}
\caption{Semantics of formulas}
\label{fig:semantics}
\end{figure*}

\subsection{Sequents and Rules}
\label{sec:sequents-and-rules}

Some proof systems have been proposed for LTL-like
logics~\cite{Bolotov+Basukoski+Grigoriev+Shangin:2006,Bolotov+Grigoriev+Shangin:2007}.%
However, the until operator is known to pose
difficulties~\cite{Masini+Vigano+Volpe:2010}. A labelled natural
deduction system was proposed in~\cite{Volpe:2010} to circumvent those
difficulties, following a long line of work on labelled deduction
systems~\cite{Gabbay:1996,Simpson:phd:1994, Basin+Caleiro+Ramos+Vigano_2009, vigano2008}. Simplifying and generalizing that system, we
propose a labelled sequent calculus that allows reasoning about until
and timing properties using the same machinery.%

A sequent is of the form $\seq{\Gamma}{\annot}{\varphi}$, where
$\Gamma$ is a list of hypotheses, $\varphi$ is a formula, and $\annot$
is a time annotation.
Time annotations are as follows, where $\ANno$ indicates no
annotation, meaning the formula is time-irrelevant~\flink{161}:
$$\annot::=\ANno\mid{t}\mid\ANintcc{t_1}{t_2}\mid\ANintco{t_1}{t_2}\mid\ANintoc{t_1}{t_2}\mid\ANintoo{t_1}{t_2}$$
Such a sequent states that $\varphi$ holds at time $\annot$ in the
context~$\Gamma$.
A hypothesis is of the form $\hyp{\varphi}{\annot}$. We write~$\varphi$ for
the annotated formula~$\hyp{\varphi}{\ANno}$.
Similar to~\cite{Volpe:2010}, \calc relies on intervals to capture the
meaning of $\OPuntilSYMB$.
The formula $\OPuntil{\varphi}{\psi}$ is true at some time $t$
whenever there exists a later time $t'$ such that $\varphi$ is true in
the interval $\ANintco{t}{t'}$ and $\psi$ is true at $t'$.

As explained below, some rules are only valid when formulas are
annotated with \emph{non-empty annotations}. %
A non-empty annotation is either of the form $\ANno$, or $t$, or an
interval as long as it is non-empty, e.g., of the form
$\ANintcc{t_1}{t_2}$ if $\OPtle{t_1}{t_2}$ holds.
Therefore, to state such rules, \calc also features sequents of the
form $\nseq{\Gamma}{\annot}$, stating that $\annot$ is a non-empty
annotation in the context $\Gamma$.

The semantics of annotated formulas is defined as follows~\flink{164}:
$$\begin{array}{c}
\begin{array}{lll}
\interpp{\run}{\pi}{\val}{\hyp{\varphi}{\ANno}} & \iff & \interp{\run}{\pi}{\Mzero}{\val}{\varphi}\\
\end{array}
\hspace{0.4in}
\begin{array}{lll}
\interpp{\run}{\pi}{\val}{\hyp{\varphi}{t}} & \iff & \interp{\run}{\pi}{\sem{t}{\val}}{\val}{\varphi}\\
\end{array}
\\
\begin{array}{lll}
\interpp{\run}{\pi}{\val}{\hyp{\varphi}{\ANintcc{t_1}{t_2}}} & \iff &
  \forall{w}.{\Mfun{\Mle{\sem{t_1}{\val}}{\Mle{w}{\sem{t_2}{\val}}}}{\interp{\run}{\pi}{w}{\val}{\varphi}}}\\
\multicolumn{3}{l}{\mbox{and similarly for the other intervals}}
\end{array}
\end{array}$$
The semantics of (1) a list of hypotheses $\Gamma$ of the form
$h_1,\dots,h_n$, (2) a sequent~$S$ of the form
$\seq{\Gamma}{\annot}{\varphi}$, and (3) a rule~$R$ with premises
$S_1,\dots,S_n$ and conclusion $S$, is defined as follows~\flink{167}:
$$\begin{array}{lll}
\interpp{\run}{\pi}{\val}{\Gamma}
& \iff
& \forall{i\in[1,n]}.{\interpp{\run}{\pi}{\val}{h_i}}
\\
\interppp{\run}{\pi}{S}
& \iff
& \forall{\val(\Gamma,\annot,\varphi)}.\:{\Mfun{\interpp{\run}{\pi}{\val}{\Gamma}}{\interpp{\run}{\pi}{\val}{\hyp{\varphi}{\annot}}}}
\\
\interppp{\run}{\pi}{R}
& \iff
& \Mfun{\interppp{\run}{\pi}{S_1}}{\Mfun{\cdots}{\Mfun{\interppp{\run}{\pi}{S_n}}{\interppp{\run}{\pi}{S}}}}
\end{array}$$
A rule $R$ is valid iff $\interppp{\run}{\pi}{R}$ for all runs $r$ and
interpretations~$\pi$.
The above semantics is formalised in Agda, and we have proved that
\calc's rules presented below are valid.

\subsection{Sequent Calculus}
\label{sec:sequent-calculus}

We now present an excerpt of \calc's rules. See
Appx.~\ref{sec:full-set-of-rules} for the full list.

\paragraph{Propositional Logic Rules}
The propositional logic rules are all standard~\flink{176}:
\[
\begin{small}
\begin{array}{c}
  \inference{\nseq{\Gamma,\Delta}{\annot_1}}
            {\seq{\Gamma,\hyp{\OPfalse}{\annot_1},\Delta}{\annot}{\varphi}}
            [$\RULEfalseL$]
  \hspace{0.2in}
  \inference{\seq{\Gamma}{\annot}{\varphi} & \seq{\Gamma}{\annot}{\psi}}
            {\seq{\Gamma}{\annot}{\OPand{\varphi}{\psi}}}
            [$\RULEandR$]
  \hspace{0.2in}
  \inference{\seq{\Gamma,\hyp{\varphi}{\annot},\hyp{\psi}{\annot},\Delta}{\annot_1}{\delta}}
            {\seq{\Gamma,\hyp{\OPand{\varphi}{\psi}}{\annot},\Delta}{\annot_1}{\delta}}
            [$\RULEandL$]
\end{array}
\end{small}
\]
\vspace{-0.05in}
\[
\begin{small}
\begin{array}{c}
  \inference{\seq{\Gamma}{\annot}{\varphi}}
            {\seq{\Gamma}{\annot}{\OPor{\varphi}{\psi}}}
            [$\RULEorRa$]
  \hspace{0.2in}
  \inference{\seq{\Gamma}{\annot}{\psi}}
            {\seq{\Gamma}{\annot}{\OPor{\varphi}{\psi}}}
            [$\RULEorRb$]
  \hspace{0.2in}
  \inference{\seq{\Gamma,\hyp{\varphi}{t},\Delta}{\annot}{\delta} & \seq{\Gamma,\hyp{\psi}{t},\Delta}{\annot}{\delta}}
            {\seq{\Gamma,\hyp{\OPor{\varphi}{\psi}}{t},\Delta}{\annot}{\delta}}
            [$\RULEorL$]
\end{array}
\end{small}
\]
\vspace{-0.05in}
\[
\begin{small}
\begin{array}{c}
  \inference{\seq{\Gamma,\hyp{\varphi}{t}}{t}{\psi}}
            {\seq{\Gamma}{t}{\OPimp{\varphi}{\psi}}}
            [$\RULEimpR$]
  \hspace{0.2in}
  \inference{\seq{\Gamma,\Delta}{t}{\varphi} & \seq{\Gamma,\hyp{\psi}{t},\Delta}{\annot}{\delta}}
            {\seq{\Gamma,\hyp{\OPimp{\varphi}{\psi}}{t},\Delta}{\annot}{\delta}}
            [$\RULEimpL$]
  \hspace{0.2in}
  \inference{}
            {\seq{\Gamma,\hyp{\varphi}{\annot}}{\annot}{\varphi}}
            [$\RULEvar$]
\end{array}
\end{small}
\]

Note that the disjunction in the $\RULEorL$ rule cannot be
annotated with an interval, since the semantics of intervals is a
$\forall$, and $\forall$ does not distribute over~$\vee$.
Similarly,  $\RULEimpR$ cannot be annotated with
intervals either.

\paragraph{Non-Empty Annotations}
\label{appx:non-empty-intervals}

Another issue arises with $\OPfalse\mbox{L}$, which would not hold without the premise
$\nseq{\Gamma,\Delta}{\annot_1}$. For example, the instance
$\seq{\hyp{\OPfalse}{\annot_1}}{\ANno}{\OPfalse}$ is semantically
false when~$\annot_1$ is an empty interval because
$\hyp{\OPfalse}{\annot_1}$ is vacuously true.
\calc features the following rules to prove that annotations are
non-empty~\flink{197}:
\[
\begin{small}
\begin{array}{c}
\inference{\seqb{\Gamma}{\OPtle{t_1}{t_2}}}
          {\nseq{\Gamma}{\ANintcc{t_1}{t_2}}}
          []
\hspace{0.1in}
\inference{\seqb{\Gamma}{\OPtlt{t_1}{t_2}}}
          {\nseq{\Gamma}{\ANintco{t_1}{t_2}}}
          []
\hspace{0.1in}
\inference{\seqb{\Gamma}{\OPtlt{t_1}{t_2}}}
          {\nseq{\Gamma}{\ANintoc{t_1}{t_2}}}
          []
\hspace{0.1in}
\inference{\seqb{\Gamma}{\OPtlt{t_1}{t}} & \seqb{\Gamma}{\OPtlt{t}{t_2}}}
          {\nseq{\Gamma}{\ANintoo{t_1}{t_2}}}
          []
\end{array}
\end{small}
\]
\vspace{-0.05in}
\[
\begin{small}
\begin{array}{c}
\inference{}
          {\nseq{\Gamma}{t}}
          []
\hspace{0.1in}
\inference{}
          {\nseq{\Gamma}{\ANno}}
          []
\end{array}
\end{small}
\]

\paragraph{Temporal Logic Rules}
We now present the temporal rules, which illustrate key aspects of our
calculus, where $\tvar$ is a ``fresh'' variable~\flink{212}:
\[
\begin{small}
\begin{array}{c@{\hspace{0.1in}}c}
  \inference{\seqb{\Gamma}{\OPtprev{t}{t_1}}
             & \seq{\Gamma}{t_1}{\varphi}}
            {\seq{\Gamma}{t}{\OPnext{\varphi}}}
            [$\RULEnextR$]
  &
    \inference{\seqb{\Gamma}{\OPtle{t}{t_1}}
               & \seq{\Gamma}{t_1}{\psi}
               & \seq{\Gamma}{\ANintco{t}{t_1}}{\varphi}}
              {\seq{\Gamma}{t}{\OPuntil{\varphi}{\psi}}}
              [$\RULEuntilR$]
\end{array}
\end{small}
\]
\vspace{-0.05in}
\[
\begin{small}
\begin{array}{c@{\hspace{0.1in}}c}
  \inference{\seq{\Gamma,\OPtprev{t}{\tvar},\hyp{\varphi}{\tvar},\Delta}{\annot}{\delta}}
            {\seq{\Gamma,\hyp{\OPnext{\varphi}}{t},\Delta}{\annot}{\delta}}
            [$\RULEnextL$]
  &
    \inference{\seq{\Gamma,\OPtle{t}{\tvar},\hyp{\psi}{\tvar},\hyp{\varphi}{\ANintco{t}{\tvar}},\Delta}{\annot}{\delta}}
              {\seq{\Gamma,\hyp{\OPuntil{\varphi}{\psi}}{t},\Delta}{\annot}{\delta}}
              [$\RULEuntilL$]
\end{array}
\end{small}
\]
The $\OPuntilSYMB${R} rule states that for $\OPuntil{\varphi}{\psi}$
to be true at time~$t$, there must exist a time~$t_1$ greater than or
equal to~$t$ at which~$\psi$ holds, and~$\varphi$ holds between~$t$
and~$t_1$.
The $\OPuntilSYMB${L} rule states that if $\OPuntil{\varphi}{\psi}$ is
true at time $t$ then there must be a time $\tvar$ (a variable)
greater than or equal to~$t$ at which $\psi$ holds, and~$\varphi$ must
be true between~$t$ and~$\tvar$.

\paragraph{Timed Rules}
The rules for the comparison and freeze operators are as follows (we
only show some of the comparison rules, as they reflect the axioms
in Fig.~\ref{fig:world-axioms})~\flink{221}:
\[
\begin{small}
\begin{array}{c}
  \inference{\seq{\Gamma,\hyp{\Msub{\varphi}{\tvar}{t}}{t},\Delta}{\annot}{\psi}}
            {\seq{\Gamma,\hyp{\OPfreeze{\tvar}{\varphi}}{t},\Delta}{\annot}{\psi}}
            [$\RULEfreezeL$]
  \hspace{0.1in}
  \inference{\seq{\Gamma}{t}{\Msub{\varphi}{\tvar}{t}}}
            {\seq{\Gamma}{t}{\OPfreeze{\tvar}{\varphi}}}
            [$\RULEfreezeR$]
\end{array}
\end{small}
\]
\vspace{-0.1in}
\[
\begin{small}
\begin{array}{c}
  \inference{}
            {\seq{\Gamma}{\annot}{\OPteq{\OPtadd{t_1}{t_2}}{\OPtadd{t_2}{t_1}}}}
            [$\RULEaddSym$]
  \hspace{0.1in}
  \inference{}
            {\seq{\Gamma,\hyp{\OPtle{t_1}{t_2}}{},\Delta}{\annot}{\OPtle{t_1}{t_2}}}
            [$\RULEvarle$]
\end{array}
\end{small}
\]
Note that the annotation on the hypothesis $\OPtle{t_1}{t_2}$ of the
$\RULEvarle$ rule could be any annotation except an empty interval,
unless~$\annot$ is also an empty interval.

Furthermore, \calc's $\RULEiccL$, $\RULEiocL$, $\RULEicoL$,
$\RULEiooL$, $\RULEiccR$, $\RULEiocR$, $\RULEicoR$, and $\RULEiooR$
rules allow instantiating intervals, such as the following ones, where
$\tvar$ is a fresh variable~\flink{230}:
\[
\begin{small}
\begin{array}{c}
\begin{array}{c@{\hspace{0.1in}}c}
\inference{\seq{\Gamma,\hyp{\varphi}{t},\Delta}{\annot}{\psi}
           & \seqb{\Gamma,\Delta}{\OPtle{t_1}{t}}
           & \seqb{\Gamma,\Delta}{\OPtle{t}{t_2}}}
          {\seq{\Gamma,\hyp{\varphi}{\ANintcc{t_1}{t_2}},\Delta}{\annot}{\psi}}
          [$\RULEiooL$]
&
\inference{\seq{\Gamma,\OPtle{t_1}{\tvar},\OPtle{\tvar}{t_2}}{\tvar}{\phi}}
          {\seq{\Gamma}{\ANintcc{t_1}{t_2}}{\phi}}
          [$\RULEiooR$]
\end{array}
\\ \\
\begin{array}{c@{\hspace{0.1in}}c}
\inference{\seq{\Gamma,\hyp{\varphi}{t},\Delta}{\annot}{\psi}
           & \seqb{\Gamma,\Delta}{\OPtle{t_1}{t}}
           & \seqb{\Gamma,\Delta}{\OPtlt{t}{t_2}}}
          {\seq{\Gamma,\hyp{\varphi}{\ANintco{t_1}{t_2}},\Delta}{\annot}{\psi}}
          [$\RULEiocL$]
&
\inference{\seq{\Gamma,\OPtle{t_1}{\tvar},\OPtlt{\tvar}{t_2}}{\tvar}{\phi}}
          {\seq{\Gamma}{\ANintco{t_1}{t_2}}{\phi}}
          [$\RULEiocR$]
\end{array}
\end{array}
\end{small}
\]

\paragraph{Induction}
\label{sec:rule-induction}
Furthermore, assuming that worlds satisfy the induction axiom in
Fig.~\ref{fig:world-axioms}, the following rule holds, assuming that
$\tvar$ does not occur free in $\Gamma$~\flink{233}:
\[
\begin{small}
\inference{\seq{\Gamma,\hyp{\OPfreeze{\tvar}{\varphi}}{\ANintco{\OPtzero}{\tvar}}}{\tvar}{\varphi}}
          {\seq{\Gamma}{\tvar}{\varphi}}
          [$\RULEind$]
\end{small}
\]

Note the use of the freeze operator in the induction hypothesis (IH)
$\hyp{\OPfreeze{\tvar}{\varphi}}{\ANintco{\OPtzero}{\tvar}}$. It is
necessary because not only is $\tvar$ the time at which $\varphi$ is
stated to hold, but~$\tvar$ can also occur free in~$\varphi$.
If the IH were $\hyp{\varphi}{\ANintco{\OPtzero}{\tvar}}$ (without the
freeze operator), then we would be assuming that for any time $t$ in
$\ANintco{\OPtzero}{\tvar}$, the formula $\varphi$ holds at~$t$, but
where the variable $\tvar$ in~$\varphi$ is still~$\tvar$. However, we
want the variable~$\tvar$ to be~$t$ too, which the freeze operator
achieves by binding~$\tvar$'s occurrences in~$\varphi$.

Another naive attempt at stating this induction hypothesis is
$\hyp{\OPprev{(\OPalwaysbefore{\OPfreeze{\tvar}{\varphi}})}}{\tvar}$.
However, this fails in capturing the induction hypothesis
because~$\OPprevSYMB$ requires the existence of a world
preceding~$\tvar$, which might not exist. To state the induction
hypothesis in that fashion, \calc would need to include an alternative
$\OPprevSYMB$ operator, say $\OPprevbSYMB$, with semantics
$\forall{\Mpred{w'}{w}}.(\interp{\run}{\pi}{w'}{\val}{\varphi})$.
We instead rely on \calc's intervals as shown above.
Finally, the hypothesis would have to be
$\hyp{\OPprevb{(\OPalwaysbefore{\OPfreeze{\tvar}{\varphi}})}}{\tvar}$
and not
$\hyp{\OPfreeze{\tvar}{\OPprev{(\OPalwaysbefore{\varphi})}}}{\tvar}$,
which has a different semantics.

\paragraph{Classical Rules}
\calc features a $\RULElem$ rule to make classical reasoning
 explicit without multiple conclusions~\flink{236}:
\[
\begin{small}
\inference{}
          {\seq{\Gamma}{\annot}{\OPor{\varphi}{\OPneg{\varphi}}}}
          [$\RULElem$]
\end{small}
\]

\paragraph{Derived Rules}
Using the above rules, in particular $\OPuntilSYMB${R} and
$\OPuntilSYMB${L}, right and left rules can be derived for
$\OPalwaysSYMB$ and $\OPeventuallySYMB$, and their past, bounded, and
strict versions. These rules will in particular be critical in proving
properties of real-time systems as illustrated with a simple example
in Sec.~\ref{sec:examples-of-simple-proofs}.
In all the rules below,~$\tvar$ stands for a ``fresh'' variable.
For example, the following rules hold for $\OPalwaysSYMB$, the bounded
version of~$\OPeventuallySYMB$, and $\OPstrictbeforeSYMB$~\flink{239}:
\[
\begin{small}
\begin{array}{l@{\hspace{0.1in}}l}
  \inference{\seq{\Gamma,\OPtle{t}{\tvar}}{\tvar}{\varphi}}
            {\seq{\Gamma}{t}{\OPalways{\varphi}}}
            [$\RULEalwaysR$]
  &
  \inference{\seq{\Gamma,\hyp{\varphi}{t_2},\Delta}{t}{\psi}
             & \seqb{\Gamma,\Delta}{\OPtle{t_1}{t_2}}}
            {\seq{\Gamma,\hyp{\OPalways{\varphi}}{t_1},\Delta}{t}{\psi}}
            [$\RULEalwaysL$]
\end{array}
\end{small}
\]
\begin{small}
\[
\begin{array}{l@{\hspace{0.1in}}l}
  \inference{\seq{\Gamma}{t_1}{\varphi} & \seqb{\Gamma}{\OPtlt{t_1}{t}}}
            {\seq{\Gamma}{t}{\OPstrictbefore{\varphi}}}
            [$\RULEstrictbeforeR$]
  &
  \inference{\seq{\Gamma}{t_1}{\varphi} & \seqb{\Gamma}{\OPtle{t}{t_1}} & \seqb{\Gamma}{\OPtle{t_1}{\OPtadd{t_0}{t}}}}
            {\seq{\Gamma}{t}{\OPeventuallyby{t_0}{\varphi}}}
            [$\RULEeventuallybyR$]
  \\ \\
  \inference{\seq{\Gamma,\hyp{\varphi}{\tvar},\OPtlt{\tvar}{t_1},\Delta}{t}{\psi}}
            {\seq{\Gamma,\hyp{\OPstrictbefore{\varphi}}{t_1},\Delta}{t}{\psi}}
            [$\RULEstrictbeforeL$]
  &
  \inference{\seq{\Gamma,\hyp{\varphi}{\tvar},\OPtle{t_1}{\tvar},\OPtle{\tvar}{\OPtadd{t_0}{t_1}},\Delta}{t}{\psi}}
            {\seq{\Gamma,\hyp{\OPeventuallyby{t_0}{\varphi}}{t_1},\Delta}{t}{\psi}}
            [$\RULEeventuallybyL$]
\end{array}
\]
\end{small}

\paragraph{Soundness}
\begin{theorem}
 \calc is sound w.r.t.\ the semantics presented in
 Sec.~\ref{sec:semantics} and Sec.~\ref{sec:sequents-and-rules}, i.e.,
 all the rules presented above are valid.
\end{theorem}

While most rules can be proved valid constructively in the metatheory
(in Agda), some require classical logic such as $\RULElem$ and
$\RULEalwaysL$.

\subsection{Example}
\label{sec:examples-of-simple-proofs}

Sec.~\ref{sec:application-simple} introduced a simple forwarding
example, where agents forward the messages they receive in a
synchronous network, guaranteeing that any message sent to a single
agent is received by all agents by some time bound. As sketched below,
\calc allows proving the following property (see
Appx.~\ref{sec:forward-proof} for full details):
\begin{lemma}
\label{lem:forwarding-property}
The following holds in all models~\flink{252}:
$\seq{\hypp{eq.~\eqref{equ:simple-example-hyp1}}{t},\hypp{eq.~\eqref{equ:simple-example-hyp2}}{t}}{t}{eq.~\eqref{equ:simple-example-concl}}$.
\end{lemma}

\begin{proof}
Assume $\hypp{eq.~\eqref{equ:simple-example-hyp1}}{t}$ and
$\hypp{eq.~\eqref{equ:simple-example-hyp2}}{t}$, agents $a$, $b$, and
$c$, and data $d$, and prove that
$\OPalways{(\OPimp{\ATsend{a}{d}{[b]}}{\OPeventuallyby{2T}{\ATreceive{c}{d}{b}}})}$
holds at $t$.
By $\RULEalwaysR$, it is enough to prove that
$\OPeventuallyby{2T}{\ATreceive{c}{d}{b}}$ holds at some time $\tvar$,
assuming $\OPtle{t}{\tvar}$ and $\hyp{\ATsend{a}{d}{[b]}}{\tvar}$.
In particular, from $\hyp{\ATsend{a}{d}{[b]}}{\tvar}$ and
$\hypp{eq.~\eqref{equ:simple-example-hyp1}}{t}$, using $\RULEalwaysL$,
we get to further assume that
$\hyp{\OPeventuallyby{T}{(\OPimp{\OPin{b}{[b]}}{\ATreceive{b}{d}{a}})}}{\tvar}$.
Now by $\RULEeventuallybyL$, we can refine this assumption to
$\hyp{\ATreceive{b}{d}{a}}{\tvar'}$ for some $\tvar'$ such that
$\OPtle{\tvar}{\tvar'}$ and $\OPtle{\tvar'}{\OPtadd{T}{\tvar}}$.
From $\hyp{\ATreceive{b}{d}{a}}{\tvar'}$ and
$\hypp{eq.~\eqref{equ:simple-example-hyp2}}{t}$, using $\RULEalwaysL$,
we get to further assume $\hyp{\ATsend{b}{d}{[c]}}{\tvar'}$.
Once again, using $\hypp{eq.~\eqref{equ:simple-example-hyp1}}{t}$ and
$\RULEalwaysL$ on that hypothesis, we get to assume
$\hyp{\OPeventuallyby{T}{(\OPimp{\OPin{c}{[c]}}{\ATreceive{c}{d}{b}})}}{\tvar'}$,
and using $\RULEeventuallybyL$, we can refine this assumption to
$\hyp{\ATreceive{c}{d}{b}}{\tvar''}$ for some $\tvar''$ such that
$\OPtle{\tvar'}{\tvar''}$ and $\OPtle{\tvar''}{\OPtadd{T}{\tvar'}}$.
Finally, we conclude using $\RULEeventuallybyR$, from the fact that
$\OPtle{\tvar}{\tvar''}$ and $\OPtle{\tvar''}{\OPtadd{2T}{\tvar}}$.
\end{proof}

\section{Real-Time Distributed Systems}
\label{sec:modular-pistis}

\subsection{The Pistis Real-Time Reliable Broadcast Protocol}

Our main use case is the Pistis real-time Byzantine-resilient reliable
broadcast
protocol~\cite{Kozhaya+Decouchant+Rahli+Verissimo:tpds:2021}.
A broadcast protocol allows messages sent by (correct) nodes
in a
distributed system to be eventually received (or delivered) by the
other (correct) nodes when the network conditions are ``good enough''.
We are in particular interested here in scenarios where nodes can be
faulty, where faults can range from benign crashes to more critical
adversarial attacks. This fault model is called the \emph{Byzantine}
fault model~\cite[Sec.~2.2.6]{cachin_introduction_2011}, in which
faulty nodes behave arbitrarily.
We further assume that nodes can be authenticated, usually
achieved through digital signatures.
All messages are signed by their senders,
meaning that correct nodes cannot be impersonated.
Many Byzantine-resilient broadcast protocols exist, providing
different properties. In particular, Byzantine consistent
broadcast~\cite[Sec.~3.10]{cachin_introduction_2011}
guarantees that if a correct node broadcasts a message then that
message will be delivered by all other correct nodes. This however
does not cover the scenario where the broadcasting node is faulty, in
which case only some of the other correct nodes might deliver the
message. Byzantine reliable
broadcast~\cite[Sec.~3.11]{cachin_introduction_2011} further guarantees
that a correct node delivers a message iff all correct nodes deliver
that message.
More precisely, the guarantees provided by a Byzantine reliable
broadcast protocol are:
\emph{Validity}, i.e., if a correct node broadcasts a message, then every
correct node eventually delivers that message;
\emph{No duplication}, i.e., every correct node delivers at most one
message;
\emph{Integrity}, i.e., if a correct node delivers a message with
correct sender~$i$, then that message was broadcast
by~$i$;%
\emph{Consistency}, i.e., if two correct nodes~$i$ and~$j$ deliver a
message, then those messages are the same;
and \emph{Totality}, i.e., if a correct node delivers a message then every
correct node eventually delivers a message.

Many Byzantine reliable broadcast algorithms have been developed, such
as Bracha's algorithm~\cite{Bracha:iandc:1987}, which requires
$3f{+}1$ nodes to tolerate up to $f$ Byzantine faults, and
makes use of two message transmission phases (often called \emph{echo}
and \emph{ready}) to guarantee the above properties.
More recently, algorithms have been designed that further guarantee a
timeliness
property~\cite{RT-ByzCast,Kozhaya+Decouchant+Rahli+Verissimo:tpds:2021}.
For example,
Pistis~\cite{Kozhaya+Decouchant+Rahli+Verissimo:tpds:2021} builds on
top of Bracha's algorithm, and further satisfies
\emph{Timeliness}, i.e., there exists a known bound $\Delta$ such that
  if a correct node broadcasts a message $m$ at real-time~$t$, then no
  correct node delivers~$m$ after~$t+\Delta$.

Pistis achieves timeliness thanks to a \emph{proof-of-connectivity}
(PoC) mechanism that requires nodes to become passive
when not receiving messages in a timely manner, as they are then
likely to deliver messages late.%
Timeliness would be easily achievable in a \emph{synchronous} network
where message transmission is bounded, and the bound, here called~$d$,
is known to the nodes.
Instead,~\cite{Kozhaya+Decouchant+Rahli+Verissimo:tpds:2021} assumes a
more realistic \emph{probabilistic synchronous} network, where messages can
arrive after the bound~$d$ with low probability.
Pistis would still satisfy the above properties in a completely
\emph{asynchronous} network where message transmission is unbounded. However,
frequent delays beyond~$d$ would result in the PoC mechanism making
too many nodes become passive, resulting in a system shutdown.
Probabilistic synchrony, on the other hand, guarantees that system
shutdowns are unlikely.

\begin{figure}[!t]
\begin{center}
  \hspace*{-0.4in}
  \vspace*{-0.1in}
  \includegraphics[width=0.49\textwidth]{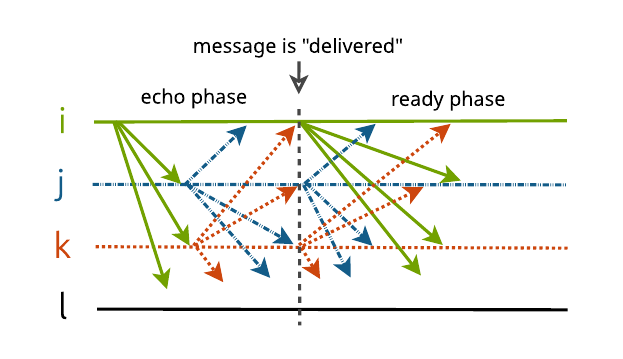}
  \includegraphics[width=0.54\textwidth]{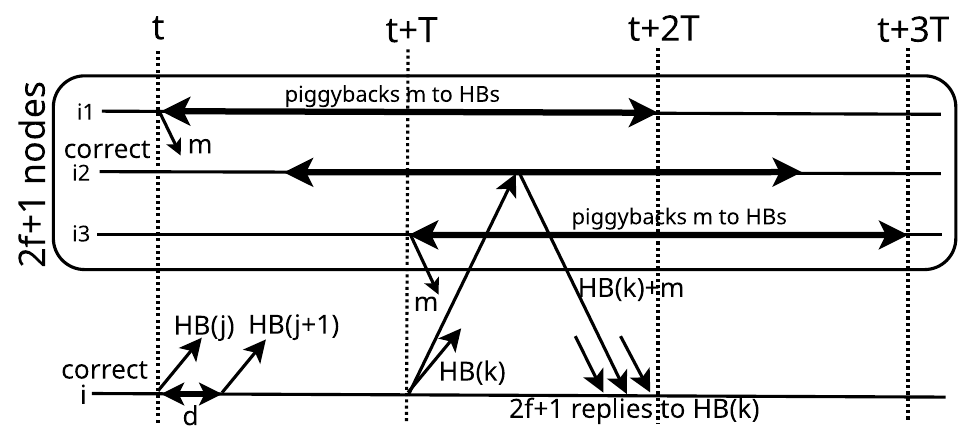}
\end{center}
\caption{Pistis: high-level overview (left) \& pulling mechanism (right)}
\label{fig:pulling-pistis}
\label{fig:high-level-pistis}
\end{figure}

Fig.~\ref{fig:high-level-pistis} illustrates the Pistis protocol with
$4$ nodes, i.e., at most $1$ faulty node. Node~$i$ broadcasts a
message by sending an echo message to all other nodes, which reply by
sending their own echo messages to all other nodes. Nodes that receive
$2f{+}1$ echo messages,\footnote{Since there are $3f{+}1$ nodes, $f$
of which might be faulty, nodes can only expect to receive messages
from $2f{+}1$ nodes.} including their own, (or a ready message)
proceed by sending a ready message to all other nodes. Any node that
has not received $2f{+}1$ echo messages (or one ready message)~$T$
units of time after it has sent its echo message becomes passive.
Similarly, any node that has not received $2f{+}1$ ready messages $2T$
units of time after it has sent its ready message becomes passive. We
expand on those~$T$ and~$2T$ bounds below.%

We now introduce Pistis in more detail, describing its main
functionalities in terms of pushing and pulling messages, leading to a
discussion on how it achieves timeliness, while
Sec.~\ref{sec:application} illustrates how to verify real-time
properties of distributed systems using \calc by proving a property
crucial for timeliness.

\paragraph{Pushing}
A node that sends a message at time~$t$ without receiving $2f{+}1$
replies by $t{+}T$ becomes passive.
Here, $T$ is a multiple of $d$, the maximum transmission delay, picked
by the system administrator depending on the quality of the network.
Assuming synchrony, it would be guaranteed that all messages
sent at~$t$ are received by correct nodes by~$t{+}d$.
However, assuming probabilistic synchrony, we can only derive the
weaker \emph{``quorum'' synchrony assumption} guaranteeing that if a
correct node sends a message at time~$t$, then a ``quorum'' of
$2f{+}1$ nodes are known to have received it by time $t{+}T$.
This property can be stated as follows, where a ``Quorum'' is
a set of $2f{+}1$ nodes, $f{+1}$ of which are therefore correct~\flink{277}:
\begin{equation}
\begin{array}{l}
\forall{p:\mbox{Node}}.
~\mbox{correct}(p)
\rightarrow
\forall{m:\mbox{Msg}}.
\forall{t:\mbox{Time}}.
~p~\mbox{sends}~m~\mbox{at}~t\rightarrow\\
\exists{Q:\mbox{Quorum}}.
\forall{q\in{Q}}.\\
\hspace*{0.2in}p~\mbox{receives the acknowledgment}~\mbox{within}~[t,t+T]
~\mbox{that}~q~\mbox{received}~m\\
\end{array}
\label{equ:p1}
\end{equation}

\paragraph{Pulling}
To guarantee timeliness despite messages arriving
late, nodes implement the above PoC mechanism, which
sends \emph{heartbeat} messages every $d$ units of time  to probe the
rest of the system and fetch (or \emph{pull}) broadcast messages. A
node that does not receive enough replies to its heartbeats becomes
passive, as it risks delivering late.
Heartbeats are tagged with unique identifiers, and given a heartbeat
sent by node~$i$, tagged with identifier~$k$, denoted $\mbox{HB}(k)$ in
Fig.~\ref{fig:pulling-pistis}, and sent at time~$t$, $2f{+}1$ replies
to $\mbox{HB}(k)$ are expected by $t{+}T$ for~$i$ to stay active and
therefore correct.
PoC guarantees that if a quorum of nodes sends a
message
within
$[t,t{+}T]$, then all correct nodes will receive that message within
$[t{+}T,t{+}2T]$, thereby preventing nodes from staying
active while not delivering:
\begin{equation}
\begin{array}{l}
\forall{m:\mbox{Msg}}.\forall{t:\mbox{Time}}.\forall{Q:\mbox{Quorum}}.
(\forall{q\in{Q}}.~q~\mbox{sends}~m~\mbox{within}~[t,t{+}T])\rightarrow\\
\forall{p:\mbox{Node}}.
~\mbox{correct}(p)
\rightarrow
p~\mbox{receives}~m~\mbox{within}~[t{+}T,t{+}2T]
\end{array}
\label{equ:p2}
\end{equation}
A simple way to guarantee eq.~\eqref{equ:p2} is for nodes to reply to
heartbeats with the messages they have sent during the last $2T$ time
interval. A $2T$ interval
is necessary because a node ($i_1$ in
Fig.~\ref{fig:pulling-pistis}) in $Q$ sending $m$ at $t$ (one end of
the $[t,t{+}T]$ interval) piggybacks~$m$ on its replies to heartbeats
during the interval $[t,t{+}2T]$, while a node ($i_3$ in
Fig.~\ref{fig:pulling-pistis}) sending~$m$ at $t{+}T$ (the other end
of the $[t,t{+}T]$ interval) piggybacks~$m$ during the interval
$[t{+}T,t{+}3T]$, thereby intersecting on the interval
$[t{+}T,t{+}2T]$.
As Fig.~\ref{fig:pulling-pistis} illustrates, these two mechanisms
(sending and replying to heartbeats) guarantee that, during the
$[t{+}T,t{+}2T]$ interval, correct (non-passive) nodes ($i$ in
Fig.~\ref{fig:pulling-pistis}) obtain messages sent by quorums of
nodes. This holds because one correct node ($i_{2}$ in
Fig.~\ref{fig:pulling-pistis}) is necessarily at the intersection of
the $2f{+}1$ nodes sending replies to $\mbox{HB}(k)$ and the $2f{+}1$
nodes piggybacking~$m$.

\begin{figure*}[!t]
\begin{small}
\[
\begin{array}{ll}
\mbox{\textit{bounded delivery:}}
& \OPalways
{\OPall
  {i}
  {\LtyAgent}
  {\OPall
    {d}
    {\LtyData}
    {\OPimp
      {\ATcorrect{i}}
      {\OPimpb
        {\ATinternal{i}{d}}
        {\hspace*{-2.1in}\OPex
          {B}
          {\LtyAgents}
          {\OPandbb
            {\OPsize{B}{2f{+}1}}
            {\hspace*{-0.7in}\OPeventuallyby
              {T}
              {\OPbefore
                {\OPall{j}{\LtyAgent}{\OPimp{\OPin{j}{B}}{\OPimp{\ATcorrect{j}}{\OPeventuallyby{T}{\ATinternal{j}{d}}}}}}}
            }
          }
        }
      }
    }
  }
}
\\
\mbox{\textit{unbounded delivery:}}
&
\OPalways
{\OPall
  {i}
  {\LtyAgent}
  {\OPall
    {d}
    {\LtyData}
    {\OPimp
      {\ATcorrect{i}}
      {\OPimpb
        {\ATinternal{i}{d}}
        {\hspace*{-2.1in}\OPex
          {B}
          {\LtyAgents}
          {\OPandbb
            {\OPsize{B}{2f{+}1}}
            {\hspace*{-0.7in}\OPall
              {j}
              {\LtyAgent}
              {\OPimp{\OPin{j}{B}}{\OPimp{\ATcorrect{j}}{\OPeventuallyby{T}{\OPbefore{\ATinternal{j}{d}}}}}}
            }
          }
        }
      }
    }
  }
}
\\
\mbox{\textit{stay faulty:}}
&
\OPall
{i}
{\LtyAgent}
{\OPalways
  {(\OPimp
    {\ATcorrect{i}}
    {\OPalwaysbefore{\ATcorrect{i}}})
  }
}
\end{array}
\]
\end{small}
\vspace{-0.2in}
\caption{Bounded and unbounded delivery in \calc}
\label{fig:bounded-deliver}
\end{figure*}

\paragraph{Properties}
The above components guarantee timeliness, since by
prop.~\ref{equ:p1}, if $p$ is correct and broadcasts a message~$m$ at
time~$t$, then by time $t_{d}\le{t{+}T}$, it receives the
acknowledgments that $m$ was received by $2f{+}1$ nodes. It can then
start delivering and sending corresponding ready messages.
Again by prop.~\ref{equ:p1}, by time $t_{d}{+}T$, $p$ receives the
acknowledgment that this ready message was received by $2f{+}1$ nodes.
We then prove by induction on time $t_{d}$ that there is a
collection~$B$ of $2f{+}1$ nodes such that all correct nodes in $B$
started delivering~$m$ (and sending corresponding ready messages)
during some interval $[t_{d'},t_{d'}{+}T]$ some time prior to
$t_{d}{+}T$.
Therefore, their PoC components will piggyback those ready messages to
heartbeats (i.e., to pull requests) during the interval
$[t_{d'}{+}T,t_{d'}{+}2T]$.
By prop.~\ref{equ:p2}, all correct nodes receive~$m$ during that
interval.

Therefore, a key point in this proof is that, as captured by
\textit{bounded delivery} in Fig.~\ref{fig:bounded-deliver}~\flink{280}, if a
correct node delivers at time~$t$, then there exists a collection~$B$
of $2f{+}1$ nodes such that some time~$t'$ prior to $t{+}T$, all
correct nodes in $B$ have delivered by $t'{+}T$.
This follows from the fact that, as captured by the \textit{unbounded
delivery} \calc formula in Fig.~\ref{fig:bounded-deliver}, if a
correct node delivers at time~$t$, then there exists a collection of
$2f{+}1$ nodes that have delivered some time prior to $t{+}T$, a
consequence of prop.~\ref{equ:p1}.
Sec.~\ref{sec:application} shows that
\textit{unbounded delivery} implies \textit{bounded delivery},
assuming that faulty nodes cannot become correct.

\subsection{Proof of Bounded Delivery}
\label{sec:application}
\label{sec:application-pistis}

We now show how to prove that \emph{bounded delivery} follows from
\emph{unbounded delivery} (see Fig.~\ref{fig:bounded-deliver}), which is key
in proving Pistis' timeliness property.

\begin{lemma}
\label{lem:unbounded-delivery-imples-bounded-delivery}
The following sequent holds in all models~\flink{289}:
\[
\seq{\hyp{\mbox{\textit{unbounded delivery}}}{\OPtzero},\hyp{\textit{stay faulty}}{\OPtzero}}
    {\OPtzero}
    {\mbox{\textit{bounded delivery}}}
\]
\end{lemma}

\begin{proof}
Using $\RULEalwaysR$, we have to prove
$\seq{\mbox{\textit{unbounded delivery}}}{\tvar}{C}$,
where $\tvar$ is a fresh variable, and $C$ is the formula under the
$\OPalwaysSYMB$ in \textit{bounded delivery}.
Using the induction rule from Sec.~\ref{sec:sequent-calculus}, we
obtain the additional assumption
$\hyp{\OPfreeze{\tvar}{C}}{\ANintco{\OPtzero}{\tvar}}$.

We then assume some correct agent~$i$ and piece of data~$d$ that
``happened'' at~$i$ at time~$\tvar$, and have to exhibit a quorum of
agents~$B$, such that at some time~$t$ before $\OPtadd{T}{\tvar}$, $d$
``happened'' at all correct agents in $B$ by time $\OPtadd{T}{t}$ (as
stated by \textit{bounded delivery}).
Instantiating \textit{unbounded delivery} to $\tvar$ using
$\RULEalwaysL$, we know that there exists a quorum of agents~$B$, such
that~$d$ ``happened'' at all correct agents in~$B$ prior to
$\OPtadd{T}{\tvar}$.
Using the classical $\RULElem$ rule, we go by
cases on whether
$\OPeventuallyby
{T}
{\OPbefore
  {\OPall{j}{\LtyAgent}{\OPimp{\OPin{j}{B}}{\OPimp{\ATcorrect{j}}{\OPeventuallyby{T}{\ATinternal{j}{d}}}}}}}$
holds at time $\tvar$, concluding directly if it does.

Let us now assume that the above formula does not hold. Using
classical reasoning, we
derive that
$\OPalwaysby
{T}
{\OPalwaysbefore
  {\OPex{j}{\LtyAgent}{\OPand{\OPin{j}{B}}{\OPand{\ATcorrect{j}}{\OPalwaysby{T}{\OPneg{\ATinternal{j}{d}}}}}}}}$
holds at~$\tvar$. Instantiating both $\OPalwaysSYMB_{\_}$ and
$\OPalwaysbeforeSYMB$ with the current time $\tvar$, we get a correct
agent~$j$ in~$B$ such that $\OPalwaysby{T}{\OPneg{\ATinternal{j}{d}}}$
holds at~$\tvar$ (note that the assumption that~$j$ is correct is
at $\tvar$).
From \textit{unbounded delivery}, we also get
that $\OPeventuallyby{T}{\OPbefore{\ATinternal{j}{d}}}$ holds at~$\tvar$.
It then follows that $\OPstrictbefore{\ATinternal{j}{d}}$ holds
at~$\tvar$ too.
Using $\RULEstrictbeforeL$, it follows that $\ATinternal{j}{d}$ holds
at $\OPtlt{\tvar'}{\tvar}$.

Using $\RULEiocL$, we instantiate the induction
hypothesis with $\tvar'$, $j$, and $d$, and obtain a quorum of
agents $B'$ such that
$\OPeventuallyby
{T}
{\OPbefore
  {\OPall{j}{\LtyAgent}{\OPimp{\OPin{j}{B'}}{\OPimp{\ATcorrect{j}}{\OPeventuallyby{T}{\ATinternal{j}{d}}}}}}}$
holds at $\tvar'$ (we then have to prove that~$j$ is correct
at~$\OPtlt{\tvar'}{\tvar}$, which holds thanks to \textit{stay
faulty}).
We conclude by instantiating the conclusion with~$B'$, and by
showing that if $\OPeventuallyby{t}{\OPbefore{\psi}}$ holds at
$\OPtle{t_1}{t_2}$ then it also holds at~$t_2$.
\end{proof}

\section{Automating Proofs in \calc{}}
\label{sec:automation}

The soundness theorem of Sec.~\ref{sec:sequent-calculus} makes the Agda formalisation a \emph{proof checker} for \calc{}:
a derivation built from valid rules yields a valid sequent.  However, proving
the validity of a rule by hand remains tedious.  For example, the semantic
proof that the bounded-eventually rule $\RULEeventuallybyR$
\[
  \inference{\seq{\Gamma}{t_1}{\varphi} & \seqb{\Gamma}{\OPtle{t}{t_1}} & \seqb{\Gamma}{\OPtle{t_1}{\OPtadd{t_0}{t}}}}
            {\seq{\Gamma}{t}{\OPeventuallyby{t_0}{\varphi}}}
            [$\RULEeventuallybyR$]
\]
is valid takes about 20 lines of Agda~\flinka{42}. These lines unfold
$\OPeventuallySYMB_{t_0}\varphi$, the freeze quantifiers and the
satisfaction relation. They also need two \emph{weakening lemmas}.
In the formalisation, variables are de~Bruijn indices, and the type of
a formula records which variables it may contain: a formula has Agda
type \acode{Form}~$V$, where~$V$ lists the types (time, agent, data,
\dots) of the variables that may occur free in it, the most recently
bound one being index~$0$. Time expressions are indexed in the same
way. Moving a formula under $k$ new binders therefore changes its type
from \acode{Form}~$V$ to \acode{Form}~$(V, u_1, \dots, u_k)$, and its
free variables must be shifted by~$k$ to keep pointing to the same
binders. We write $\uparrow_k\!e$ for this \emph{weakening} of a
formula or time expression $e$.
Weakening is left implicit in the definition of
$\OPeventuallySYMB_{t}$ in Sec.~\ref{sec:tptl-dist}, but in Agda the
two freeze binders of $\OPeventuallySYMB_{t_0}\varphi$ force its
definition to refer to $\uparrow_2\!\varphi$ and $\uparrow_2\!t_0$.
The lemmas state that weakening does not change meaning: the value of
$\uparrow_2\!t_0$ in an environment extended with two variables is the
value of~$t_0$ in the original environment, and likewise
$\uparrow_2\!\varphi$ holds in the extended model exactly when
$\varphi$ holds in the original one.
While the hand-written proof applies them explicitly, with the
automation described below, this bookkeeping disappears from such
proofs: the derivation found by the search only combines rules whose
validity was proved once, and the weakening needed to instantiate it
is handled generically, once and for all, by the instantiation
function of Sec.~\ref{sec:automation:subst}, so that
$\RULEeventuallybyR$ is proved without mentioning any weakening lemma.
In fact the rule is a short combination of rules we have already proved
valid ($\RULEfreezeR$, $\RULEuntilR$, $\RULEandR$ and $\RULEvar$).
This section describes a proof search we implemented inside Agda that
finds such combinations automatically. It also explains how we apply
the search to goals that contain \emph{meta-level unknowns}: we
abstract the unknowns to variables, run the search, and then undo the
abstraction on the derivation it finds.

To see what these are, compare two statements. The sequent $\vdash (p
\OPandSYMB q) \OPimpSYMB (q \OPandSYMB p)^{\OPtzero}$, for two
particular atomic propositions $p$ and $q$, is a concrete piece of
syntax: every formula in it is built from constructors, so a program
can inspect it and decide which rule to apply. The rule
$\RULEeventuallybyR$ above, in contrast, is stated in Agda for
\emph{all} contexts $\Gamma$, formulas $\varphi$ and time expressions
$t$, $t_0$, $t_1$. Its Agda proof is a function that takes these as
arguments, and inside its body $\Gamma$, $\varphi$ and~$t$, $t_0$,
$t_1$ are Agda variables of type context, formula and time expression.
Nothing is known about their shape. The search cannot tell whether
$\varphi$ is a conjunction or an atom, and it cannot decide whether
two such unknowns are equal. Computation on them therefore gets stuck.
These Agda-level variables must not be confused with the time
variables $\tvar$ of \calc{}, which are part of the object syntax.

Everything described here is compatible with Agda's \texttt{-{}-safe} pragma.
It uses no postulates and no \texttt{TERMINATING} pragmas, and every proof it
produces is checked by Agda's type checker.  The search itself therefore never
needs to be trusted.

\subsection{Derivations and a Proof-Producing Search}
\label{sec:automation:search}

We represent derivations as an inductive family~\flinka{46} $\Delta
\der \varphi^{a}$ whose constructors are instances of rules of the
calculus of Sec.~\ref{sec:sequent-calculus}.
Each constructor takes derivations of the premises of its rule and
returns a derivation of its conclusion. For example, here are (without
their implicit arguments) the constructors for the axiom rule,
$\RULEandR$, $\RULEimpR$ and $\RULEimpL$:
\[
\begin{array}{lll}
  \acode{ax}   & : & \varphi^{a} \in \Delta \;\to\; \Delta \der \varphi^{a}\\
  \acode{∧I}  & : & \Delta \der \varphi^{a} \;\to\; \Delta \der \psi^{a}
                  \;\to\; \Delta \der (\varphi \OPandSYMB \psi)^{a}\\
  \acode{→I}  & : & \Delta, \varphi^{t} \der \psi^{t}
                  \;\to\; \Delta \der (\varphi \OPimpSYMB \psi)^{t}\\
  \acode{→L}  & : & (\varphi \OPimpSYMB \psi)^{a} \in \Delta \;\to\;
                  \Delta \der \varphi^{a} \;\to\;
                  \Delta, \psi^{a} \der \chi^{b} \;\to\; \Delta \der \chi^{b}
\end{array}
\]
Here $\varphi^{a} \in \Delta$ is a proof that $\varphi^{a}$ is one of
the hypotheses of $\Delta$, which records its position in $\Delta$.
For instance, $\acode{→I}\;\acode{ax}$ is a derivation of $\der
\varphi \OPimpSYMB \varphi^{t}$, where $\acode{ax}$ derives
$\varphi^{t} \der \varphi^{t}$, and $\acode{→I}$ moves $\varphi$ to
the right of the turnstile.
Soundness~\flinka{49},
\[
  \acode{sound} : \Delta \der \varphi^{a} \;\to\; \Delta \models \varphi^{a},
\]
follows by induction on derivations.  Each case reuses the validity proof of
the corresponding rule, so almost no new semantic reasoning is needed.

Search is a function~\flinka{52}
\[
  \acode{search⊢} : \mathbb{N} \to (\Delta : \mathsf{Ctxt})\to
  (a : \mathsf{Ann}\,\Delta) \to (\varphi : \mathsf{Form}\,\Delta) \to
  \Maybe\,(\Delta \der \varphi^{a})
\]
defined by recursion on its fuel argument $n$. At every step, i.e., on
the goal and on every subgoal it creates, it first tries to close the
goal with the axiom rule. If that fails and $n > 0$, it applies the
right rule for the goal's main connective and, if that fails, tries
the left rules on each hypothesis in turn. Whichever rule it applies,
it then calls itself with fuel $n - 1$ on each premise of that rule.
For example, $\RULEandR$ on $\Gamma \vdash \varphi \OPandSYMB
\psi^{a}$ yields the two subgoals $\Gamma \vdash \varphi^{a}$ and
$\Gamma \vdash \psi^{a}$, while $\RULEandL$ on a hypothesis $(\varphi
\OPandSYMB \psi)^{a}$ yields one subgoal whose context is extended
with~$\varphi^{a}$ and~$\psi^{a}$. When the fuel runs out, only the
axiom rule is tried, so $n$ bounds the depth of the derivations found.
Two rules need a witness: $\RULEnextR$ needs a time $t_1$ with $t
\mathrel{\OPtprevSYMB} t_1$, and $\RULEuntilR$ needs a time $t_1$ with
$t \OPtleSYMB t_1$. The search takes these witnesses from the
comparison hypotheses in the context. Since $\acode{search⊢}$ is total
and defined by structural recursion, Agda can run it during type
checking. To turn its result into a proof, we use the standard-library
function
\[
  \acode{from-just} : (x : \Maybe\,A) \to \acode{From-just}\;x
\]
whose return type is computed from its argument:
$\acode{From-just}\,(\Just\,d) = A$ and $\acode{From-just}\,\Nothing = \top$.
Hence, to prove a goal $\Delta \der \varphi^{a}$, it suffices to write
$\acode{from-just}\,(\acode{search⊢}\;n\;\Delta\;a\;\varphi)$.  When
type-checking this term, Agda runs the search.  If it returns $\Just\,d$, the
term has type $\Delta \der \varphi^{a}$ and evaluates to $d$.  If it returns
$\Nothing$, the term has type $\top$, which does not match the expected type
$\Delta \der \varphi^{a}$, so Agda rejects the definition with a type error.

\subsection{Meta-Level Unknowns and Abstraction}
\label{sec:automation:unknowns}

The search compares formulas and time expressions for equality, e.g.,
to close a goal with a hypothesis using the $\RULEvar$ rule. On
concrete syntax these comparisons compute, but get stuck on Agda
variables. Consider the goal $\Gamma \vdash \varphi \OPimpSYMB
\varphi^{t}$, where $\Gamma$, $\varphi$ and $t$ are Agda variables.
The search first applies $\RULEimpR$, which produces the subgoal
$\Gamma, \varphi^{t} \vdash \varphi^{t}$. It then tries the axiom
rule, i.e., it checks whether $\varphi^{t}$ occurs in the context. To
do so, it compares the goal with each hypothesis using the decision
procedure for equality of formulas (\acode{Form-dec}~\flinka{55}) and
of time expressions. These deciders proceed by pattern matching on the
syntax of their arguments. Since $\varphi$ and $t$ are Agda variables,
whose constructors are unknown, the deciders cannot reduce, even when
comparing $\varphi$ with itself. Normalisation of the search therefore
gets stuck. The context $\Gamma$, also an Agda variable, causes the
same problem, since the search cannot even enumerate its hypotheses.
These are exactly the goals that arise when proving a rule valid.

To support proof search, we extend the syntax of
Sec.~\ref{sec:tptl-dist} with two new kinds of \emph{object-level
variables}, indexed by natural numbers: \emph{formula variables}
$\fvar{k}$, a new constructor of formulas, and \emph{time-expression
variables} $\xvar{k}$, a new constructor of time expressions. They are
distinct from the time variables $\tvar$ bound by the freeze
quantifier: they are never bound, and only serve as placeholders for
meta-level unknowns. Their semantics is deliberately trivial: both are
interpreted as constants, a formula variable as falsity and a
time-expression variable as the initial time point. As explained
below, this is enough, because the variables are always instantiated
before soundness is used. We then \emph{abstract} a goal by replacing
each meta-level unknown with a fresh variable, using the same variable
for syntactically equal unknowns. On abstracted goals all comparisons
compute, and the search applies. For instance, our example $\Gamma
\vdash \varphi \OPimpSYMB \varphi^{t}$ becomes $\vdash {\fvar{0}
  \OPimpSYMB \fvar{0}}^{\xvar{0}}$ (the context $\Gamma$ is dealt with
separately, see Sec.~\ref{sec:automation:subst}). After $\RULEimpR$,
the subgoal is ${\fvar{0}}^{\xvar{0}} \vdash {\fvar{0}}^{\xvar{0}}$.
Comparing the goal with the hypothesis now amounts to comparing the
indices $0$ and $0$ of the two variables, which \acode{Form-dec}
decides by computation, so the axiom rule applies and the search
returns the derivation.

Because of this trivial semantics, substitution does not preserve
\emph{validity}. For example, the sequent ${\fvar{0}}^{\xvar{0}}
\vdash \OPfalse^{\xvar{0}}$ is valid, since $\fvar{0}$ is interpreted
as falsity, but its instance $\varphi^{t} \vdash \OPfalse^{t}$ is not.
Substitution does, however, preserve \emph{derivability}: the rules of
the calculus are schematic and none of them depends on the meaning of
a variable, so instantiating the variables in a derivation yields a
derivation. Note that ${\fvar{0}}^{\xvar{0}} \vdash
\OPfalse^{\xvar{0}}$ is not derivable. We therefore undo the
abstraction on derivations rather than on valid sequents, using a
substitution lemma on derivations, \acode{plug-⊢}~\flinka{58}
(Sec.~\ref{sec:automation:subst}). It instantiates the formula and
time-expression variables of a derivation of an abstract sequent and
plugs the result into an arbitrary context $\Gamma$. The instance is
then valid by soundness. For the same reason, the meaning of the
variables plays no role in the soundness proof, which is why constants
suffice.

This gives a proof in three steps: (i) abstract the goal, (ii) run
\acode{search⊢} on the abstract goal, and (iii) instantiate the derivation it
finds with \acode{plug-⊢}.  For our example, (i) turns
$\Gamma \vdash \varphi \OPimpSYMB \varphi^{t}$ into
$\vdash {\fvar{0} \OPimpSYMB \fvar{0}}^{\xvar{0}}$, (ii) finds a derivation
$d : {} \der {\fvar{0} \OPimpSYMB \fvar{0}}^{\xvar{0}}$ ($\RULEimpR$ followed by the
axiom rule), and (iii) turns $d$ into a derivation of
$\Gamma \der \varphi \OPimpSYMB \varphi^{t}$, by replacing $\fvar{0}$ by $\varphi$
and $\xvar{0}$ by $t$, and by adding $\Gamma$ to the (empty) context of the
sequent.
Step (ii) is always automatic.  In Method~1
(Sec.~\ref{sec:automation:subst}), steps (i) and (iii) are written by hand.
In Method~2 (Sec.~\ref{sec:automation:reflection}), a reflection macro
performs them as well.

\subsection{Method 1: Substitution}
\label{sec:automation:subst}

The key lemma, \acode{plug-⊢}, states that derivations of closed
sequents can be instantiated and \emph{plugged} into an arbitrary
context. Let $\Delta$ be a closed list of hypotheses whose only
unknowns are object variables, and let $\Gamma$ be an arbitrary
context. Let $\sigma$ map formula variables to formulas over the
variables of $\Gamma$, and let $\rho$ do the same for time-expression
variables. We define an instantiation $\instt{\cdot}$, which replaces
each $\fvar{i}$ by $\sigma(i)$ and each $\xvar{i}$ by $\rho(i)$, and
prove
\begin{equation}
  \label{eq:plug}
  \acode{plug-⊢}\;\Gamma\;\sigma\;\rho :\;
  \Delta \der \psi^{b}
  \;\to\;
  \Gamma, \instt{\Delta} \der \instt{\psi^{b}}
\end{equation}
by induction on derivations, where $\instt{\psi^{b}}$ instantiates
both the formula $\psi$ and its annotation $b$. Here $\Gamma,
\instt{\Delta}$ instantiates the hypotheses of $\Delta$ and appends
them to $\Gamma$, while leaving $\Gamma$'s own hypotheses untouched.

The derivation given to \acode{plug-⊢} is not written out by hand: it is
the expression $\acode{from-just}\,(\acode{search⊢}\;\dots)$ of
Sec.~\ref{sec:automation:search}, which computes it.  For example, Method~1
proves $\Gamma \der \varphi \OPimpSYMB \varphi^{t}$ with
\[
  \acode{plug-⊢}\;\Gamma\;(0 \mapsto \varphi)\;(0 \mapsto t)\;
  (\acode{from-just}\,(\acode{search⊢}\;10\;[\,]\;\xvar{0}\;(\fvar{0} \OPimpSYMB \fvar{0}))).
\]
The last argument evaluates to the derivation $\acode{→I}\;\acode{ax}$
of $\der {\fvar{0} \OPimpSYMB \fvar{0}}^{\xvar{0}}$ ($\RULEimpR$
applied to an axiom). Agda type-checks this term in two steps. First,
the type of $\acode{from-just}\,x$ depends on whether $x$ is a
$\Just$, so Agda evaluates the search until it returns
$\Just\,(\acode{→I}\;\dots)$. The derivation inside is the result of a
computation, whose type is guaranteed by the type of \acode{search⊢},
so it is not type-checked. Second, by~\eqref{eq:plug}, the type of the
whole term is computed from the sequent $\der {\fvar{0} \OPimpSYMB
  \fvar{0}}^{\xvar{0}}$, $\sigma$ and $\rho$ alone, by replacing
$\fvar{0}$ by $\varphi$ and $\xvar{0}$ by $t$, which gives $\Gamma
\der {\varphi \OPimpSYMB \varphi}^{t}$. Type-checking an application
of \acode{plug-⊢} does not run it, so \acode{plug-⊢} traverses the
derivation only when the proof itself is evaluated.

The interesting cases of the proof of~\eqref{eq:plug} are
$\RULEfreezeR$ and $\RULEuntilR$, the two rules that involve a new
time variable: $\RULEfreezeR$ eliminates the variable bound by the
freeze quantifier by substituting a time expression for it, and
applying $\RULEiooR$ to the third premise of $\RULEuntilR$ extends the
context with a fresh time variable (standing for the intermediate time
points between the two annotations). In the rest of this paragraph we
drop the subscript and write $\llparenthesis\cdot\rrparenthesis$ for
$\instt{\cdot}$, and $\llparenthesis\cdot\rrparenthesis^{+}$ for the
instantiation used under one more time variable, which replaces each
$\fvar{i}$ by $\uparrow_1\!\sigma(i)$ and each $\xvar{i}$ by
$\uparrow_1\!\rho(i)$, so that their free time variables skip the new
one. These cases rely on two commutation lemmas. First, instantiation
commutes with weakening:
$\llparenthesis\uparrow_1\!\varphi\rrparenthesis^{+} =
\uparrow_1\!\llparenthesis\varphi\rrparenthesis$. Second, it commutes
with the substitution of a time expression for a bound time variable:
\[
  \llparenthesis\Msub{\varphi}{\tvar}{t}\rrparenthesis =
  \Msub{\llparenthesis\varphi\rrparenthesis^{+}}{\tvar}{\llparenthesis t\rrparenthesis}.
\]
For example, $\RULEfreezeR$ concludes $\Delta \der (\OPfreeze{\tvar}{\varphi})^{t}$ from
$\Delta \der \Msub{\varphi}{\tvar}{t}^{t}$.  By induction, the premise yields
a derivation of
$\Gamma,\llparenthesis\Delta\rrparenthesis \der
\llparenthesis\Msub{\varphi}{\tvar}{t}\rrparenthesis^{\llparenthesis t\rrparenthesis}$,
which by the second lemma is
$\Gamma,\llparenthesis\Delta\rrparenthesis \der
\Msub{\llparenthesis\varphi\rrparenthesis^{+}}{\tvar}{\llparenthesis t\rrparenthesis}^{\llparenthesis t\rrparenthesis}$.
Applying $\RULEfreezeR$ then gives
$\Gamma,\llparenthesis\Delta\rrparenthesis \der
(\OPfreeze{\tvar}{\llparenthesis\varphi\rrparenthesis^{+}})^{\llparenthesis t\rrparenthesis}$,
which is the instance $\llparenthesis(\OPfreeze{\tvar}{\varphi})^{t}\rrparenthesis$ of the
conclusion.  Similarly, since the third premise of $\RULEuntilR$ adds a fresh time
variable to the context, it mentions the weakened formula
$\uparrow_1\!\varphi$, whose instance is matched with
$\uparrow_1\!\llparenthesis\varphi\rrparenthesis$ by the first lemma.  Both
lemmas are proved by induction over the whole syntax without function
extensionality.

One design decision is essential. The instantiation traverses formulas
and contexts, and on its way to an occurrence of $\fvar{i}$ it may
pass time variables that are not in $\Gamma$: variables bound by
freeze quantifiers in the formula, or added to the context by $\Delta$
or by the third premise of $\RULEuntilR$. If it has passed $k$ such
variables, $\fvar{i}$ is replaced by $\uparrow_k\!\sigma(i)$, so that
the free time variables of $\sigma(i)$ skip these $k$ variables (and
similarly for $\xvar{i}$ and $\rho$; going from
$\llparenthesis\cdot\rrparenthesis$ to
$\llparenthesis\cdot\rrparenthesis^{+}$ above increases $k$ by one).
Crucially, the weakening is \emph{accumulated} during the traversal
and applied once, as the single weakening $\uparrow_k$, rather than as
$k$ successive weakenings
$\uparrow_1(\cdots(\uparrow_1\!\sigma(i))\cdots)$. The two are equal,
but only propositionally, not by computation. Derived operators such
as $\OPeventuallySYMB_{t_0}\varphi$ are defined using exactly these
accumulated weakenings ($\uparrow_2\!\varphi$ and $\uparrow_2\!t_0$
under its two freeze binders). Using accumulated weakenings, the
instantiation of the abstract goal
$\OPeventuallySYMB_{\xvar{0}}\fvar{0}$ is therefore
\emph{definitionally} $\OPeventuallySYMB_{t_0}\varphi$. As a result,
eq.~\eqref{eq:plug} can be applied to the original goal without any
transport. Without this design, every binder in a derived operator
would require an explicit rewriting step, which is precisely the kind
of reasoning the search is meant to avoid.

Finally, a rule whose premises share the context of its conclusion follows
from a single sequent in which the premises have become hypotheses~\flinka{64}:
\begin{equation}
  \label{eq:hyps}
  \acode{hyps→rule} :\;
  \models\big(\Gamma, \varphi_1^{a_1}, \dots, \varphi_n^{a_n} \vdash \psi^{b}\big)
  \;\to\;
  \models
  \frac{\Gamma \vdash \varphi_1^{a_1}\quad\cdots\quad\Gamma \vdash \varphi_n^{a_n}}
       {\Gamma \vdash \psi^{b}}
\end{equation}
This follows by cut and weakening at the level of the semantics.

We can now prove that $\RULEeventuallybyR$ is valid, i.e., that it
holds for an arbitrary context~$\Gamma$, formula $\varphi$ and time
expressions $t_0$, $t$ and $t_1$. Combining the pieces, a proof
is~\flinka{69}
\begin{gather*}
  \acode{hyps→rule}\,\big(\acode{sound}\,(\acode{plug-⊢}\;\Gamma\;\sigma\;\rho\;d)\big)
  \\
  \text{with}\quad
  \sigma = [0 \mapsto \varphi],\;
  \rho = [0 \mapsto t_0, 1 \mapsto t, 2 \mapsto t_1],
\end{gather*}
where $d = \acode{from-just}\,(\acode{search⊢}\;10\;\Delta\;\xvar{1}\;
\OPeventuallySYMB_{\xvar{0}}\fvar{0})$ is the derivation automatically
found by the search with fuel $10$, and $\Delta =
\xvar{1}\OPtleSYMB\xvar{2},\;
\xvar{2}\OPtleSYMB\xvar{1}\OPtaddSYMB\xvar{0},\;{\fvar{0}}^{\xvar{2}}$.

In Method~1, the user writes this proof of $\RULEeventuallybyR$ by hand: the
abstract sequent $\Delta \vdash {\OPeventuallySYMB_{\xvar{0}}\fvar{0}}^{\xvar{1}}$
(step (i)), the substitutions $\sigma$ and $\rho$, and the applications of
\acode{hyps→rule} and \acode{plug-⊢} (step (iii)).  Only $d$ is found
automatically.  Agda checks that the instantiation matches the rule.

\subsection{Method 2: Reflection}
\label{sec:automation:reflection}

Method~2 automates the two manual steps of Method~1 with a single
macro \acode{prove}, written with Agda's reflection API. The macro is
written in place of a proof, and accepts three kinds of statements
about a sequent $S$: that $S$ is derivable ($\Delta \der S$), that $S$
is valid ($\models S$), and that a rule $S_1,\dots,S_n \,/\, S$ is
valid ($\models S_1,\dots,S_n \,/\, S$), where the premises share the
context of the conclusion. All three reduce to the same problem,
finding a derivation of a single closed sequent. For the first this is
immediate. The second follows from a derivation of $S$ by soundness.
For the third, eq.~\eqref{eq:hyps} reduces the validity of the rule to
that of $S$ with the premises added as hypotheses. The macro proceeds
as follows.
\begin{enumerate}
\item It obtains, as a quoted Agda term, the statement to be proved,
  i.e., the type expected in the Agda goal where the \acode{prove}
  tactic is written (for instance $M \models
  \RULEeventuallybyR(\Gamma, t_0, t, t_1, \varphi)$ below). It
  extracts from it the sequent $S$ (and the premises of a rule), and
  computes an abstract version $\Delta \der \psi^{b}$ of $S$
  (step~(i)), of the shape required by eq.~\eqref{eq:plug}. The
  context of $S$ is split into a part $\Gamma$, typically an Agda
  variable (or empty), and the concrete hypotheses written after it,
  such as $t \mathrel{\OPtprevSYMB} t_1$ and $A^{t_1}$ in $\Gamma, t
  \mathrel{\OPtprevSYMB} t_1, A^{t_1} \vdash \OPnextSYMB A^{t}$, where
  $t$ and $t_1$ are arbitrary time expressions. The closed context
  $\Delta$ consists of these hypotheses and, for a rule, of the
  right-hand sides $\varphi_i^{a_i}$ of its premises $\Gamma \vdash
  \varphi_i^{a_i}$, as in eq.~\eqref{eq:hyps}. In~$\Delta$ and in the
  right-hand side $\psi^{b}$ of $S$, meta-level unknowns
  (Sec.~\ref{sec:automation:unknowns}) are abstracted: formulas that
  are Agda variables, such as $A$, become formula variables
  $\fvar{k}$, and time expressions that are Agda variables, such as
  $t$ and $t_1$, become time-expression variables $\xvar{k}$, using
  the same variable for syntactically equal subterms. In the ``next''
  example above, $\Delta \der \psi^{b}$ is $\xvar{0}
  \mathrel{\OPtprevSYMB} \xvar{1}, {\fvar{0}}^{\xvar{1}} \der
  \OPnextSYMB {\fvar{0}}^{\xvar{0}}$.
\item It normalises \acode{search⊢} on $\Delta \der \psi^{b}$ (step (ii)).
\item It builds the term of Sec.~\ref{sec:automation:subst} (step (iii)):
  \acode{plug-⊢} is applied to $\Gamma$ (the part of the context of $S$ left
  after removing its concrete hypotheses in item~1), to the derivation
  $\acode{from-just}\,(\acode{search⊢}\;n\;\Delta\;\dots)$ exactly as in
  Method~1, and to~$\sigma$ and~$\rho$, which map the variables back to the original
  subterms.  Its type, $\Gamma, \instt{\Delta} \der \instt{\psi^{b}}$, is
  $S$ again (with the premises as hypotheses, for a rule).  For a validity
  statement it then applies \acode{sound}, and for a rule also
  \acode{hyps→rule}.
\end{enumerate}
Agda type-checks the resulting term against the original statement.  The whole
proof of the $\RULEeventuallybyR$ rule becomes~\flinka{73}:
\begin{align*}
  &\acode{rule◇↓R-sat} : \forall\,M\,\Gamma\,t_0\,t\,t_1\,\varphi.\;
  M \models \RULEeventuallybyR(\Gamma, t_0, t, t_1, \varphi)
  \\
  &\acode{rule◇↓R-sat}\;\dots = \acode{prove}\;10 ,
\end{align*}
and a sequent over arbitrary Agda formulas $A$, $B$ and $C$, and time
expression~$t$, is proved the
same way~\flinka{77}:
\begin{align*}
  &\acode{→-trans} : \forall\,M\,A\,B\,C\,t.\;
  M \models \big(\vdash (A \OPimpSYMB B) \OPimpSYMB (B \OPimpSYMB C) \OPimpSYMB A \OPimpSYMB C^{t}\big)
  \\
  &\acode{→-trans}\;\dots = \acode{prove}\;10 .
\end{align*}
The macro itself need not be trusted: if it abstracted the statement incorrectly,
the term would not type-check.

\paragraph{Why not substitute in the derivation term?}
With reflection, the macro has both the derivation $d$ found by the search
and the original subterms in hand as quoted terms.  A shortcut therefore
suggests itself for step (iii): replace the variables by the original
subterms directly in the quoted $d$, and let Agda check the result.  Since
the rules are schematic, each step of $d$ remains an instance of the same
rule, and this would avoid \acode{plug-⊢} and its meta-theory altogether.
This shortcut succeeds for propositional reasoning,
e.g., for $\Gamma \vdash \varphi \OPimpSYMB \varphi^{t}$, but not through freeze
quantifiers.  In $d$, the $\RULEfreezeR$ steps for $\OPeventuallySYMB_{\xvar{0}}\fvar{0}$
mention $\Msub{(\uparrow_1\!\fvar{0})}{\tvar}{t}$, which computes to
$\fvar{0}$.  Once $\fvar{0}$ is replaced by $\varphi$, $\RULEfreezeR$ expects
$\Msub{(\uparrow_1\!\varphi)}{\tvar}{t}$, which is only propositionally equal
to $\varphi$, so the term is rejected.  \acode{plug-⊢} handles this once and
for all with the commutation lemmas of Sec.~\ref{sec:automation:subst}.

\subsection{Comparison and Evaluation}
\label{sec:automation:comparison}
\label{sec:automation:experiments}

Both methods follow the same three steps and differ only in who performs
steps (i) and (iii): the user or the macro.  Both are proof-producing.  The
search, the macro and the abstraction need not be trusted, and everything
is compatible with Agda's \texttt{-{}-safe} pragma.

We evaluated both methods on the goals of
Table~\ref{tab:automation:experiments}.
While we only present a subset of those goals here, see \myurl{\VGITe}
for the full set of examples proved automatically.
Each goal is proved twice:
with Method~1, i.e., by writing the abstraction by hand and applying
\acode{plug-⊢}, and with the macro \acode{prove} (Method~2), always
with fuel 10. The upper part of the table consists of sequents over
arbitrary formulas and time expressions, some of them with concrete
hypotheses. The lower part consists of (derived) rules of the calculus
over an arbitrary context and arbitrary time expressions. Their
validity was previously proved by hand, at the level of the semantics,
and is now obtained with \acode{prove}. For $\RULEeventuallybyR$ this
replaces about 20 lines of semantic reasoning, or a 15-line
hand-written derivation, with a single line.

To quantify the search, we use an instrumented copy of \acode{search⊢}
that counts its recursive calls, i.e., the nodes of the search tree it
explores before returning, and we count the rule applications in the
derivation it returns. We also report the smallest fuel for which each
goal is solved. The times are those Agda spends type-checking each
definition, measured with Agda~2.7.0.1 and the standard library~2.1.1
using \texttt{-{}-profile=definitions} (median of ten runs).
All experiments ran in a Linux container (Ubuntu 24.04) with 4 cores
of an Intel Xeon processor at 2.10\,GHz and 15\,GB of RAM, using the
Agda binary packaged by Ubuntu. Absolute times will differ on other
machines, but the ratios between the two methods and between goals are
what matters here.

\begin{table}[t]
\centering
\caption{Goals solved automatically.  $A,B,C,D,A_i$ are arbitrary formulas,
$\Gamma$ an arbitrary context, and $t,t_0,t_1$ arbitrary time expressions.
Rules are given by their premises and conclusion.
$(A_i\OPimpSYMB A_{i+1})_{i<k}\OPimpSYMB\varphi$ stands for
$(A_0\OPimpSYMB A_1)\OPimpSYMB\cdots\OPimpSYMB(A_{k-1}\OPimpSYMB A_k)\OPimpSYMB\varphi$.  \emph{Fuel}: smallest
fuel that solves the goal.  \emph{Nodes}: search-tree nodes explored with
fuel 10.  \emph{Size}: rule applications in the derivation found with
fuel 10.  \emph{Macro} and \emph{Hand}: type-checking time (ms) with
Method~2 and Method~1, respectively.}
\label{tab:automation:experiments}
\small
\setlength{\tabcolsep}{2.4pt}
\begin{tabular}{@{}llrrrrr@{}}
\toprule
\textbf{Name} & \textbf{Goal} & \textbf{Fuel} & \textbf{Nodes} & \textbf{Size} & \textbf{Macro} & \textbf{Hand} \\
\midrule
\acode{id}       & $\vdash A \OPimpSYMB A^{t}$                                     & 1 &  2 &  2 & 87 & 15 \\
\acode{∧}        & $\vdash (A\OPandSYMB B)\OPimpSYMB B\OPandSYMB A^{t}$                      & 3 &  6 &  6 & 161 & 20 \\
\acode{→}        & $\vdash (A\OPimpSYMB B)\OPimpSYMB(B\OPimpSYMB C)\OPimpSYMB A\OPimpSYMB C^{t}$               & 5 & 29 & 28 & 272 & 28 \\
\acode{chain}$_3$ & $\vdash (A_i\OPimpSYMB A_{i+1})_{i<3}\OPimpSYMB A_0\OPimpSYMB A_3^{t}$ & 6 & 51 & 41 & 434 & 40 \\
\acode{chain}$_4$ & $\vdash (A_i\OPimpSYMB A_{i+1})_{i<4}\OPimpSYMB A_0\OPimpSYMB A_4^{t}$ & 8 & 60 & 30 & 655 & 47 \\
\acode{chain}$_5$ & $\vdash (A_i\OPimpSYMB A_{i+1})_{i<5}\OPimpSYMB A_0\OPimpSYMB A_5^{t}$ & 9 & 87 & 17 & 1408 & 57 \\
\acode{chain}$_6$ & $\vdash (A_i\OPimpSYMB A_{i+1})_{i<6}\OPimpSYMB A_0\OPimpSYMB A_6^{t}$ & 10 & 175 & 20 & 1178 & 82 \\
\acode{∨}        & $\vdash (A\OPorSYMB B)\OPimpSYMB B\OPorSYMB A^{t}$                        & 3 & 12 &  6 & 156 & 20 \\
\acode{distr}    & $\begin{array}[t]{@{}l@{}}\vdash (A\OPorSYMB B)\OPandSYMB(C\OPorSYMB D)\OPimpSYMB{}\\ \quad(A\OPandSYMB C\OPorSYMB A\OPandSYMB D)\OPorSYMB(B\OPandSYMB C\OPorSYMB B\OPandSYMB D)^{t}\end{array}$ & 7 & 620 & 25 & 1266 & 248 \\
\acode{ctx}      & $(A\OPandSYMB B)^{t} \vdash B\OPorSYMB\OPfalse^{t}$                    & 2 &  3 &  3 & 260 & 20 \\
\acode{mixed}    & $\vdash (\OPnextSYMB A \OPandSYMB \fvar{0}) \OPimpSYMB (\fvar{0}\OPandSYMB\OPnextSYMB A)^{t}$ & 3 & 6 & 6 & 212 & 20 \\
\acode{next}     & $t \mathrel{\OPtprevSYMB} t_1,\; A^{t_1} \vdash \OPnextSYMB A^{t}$            & 1 &  2 &  3 & 253 & 22 \\
\acode{event.} & $t \OPtleSYMB t_1,\; B^{t_1} \vdash \OPeventuallySYMB B^{t}$   & 2 &  3 &  4 & 375 & 24 \\
\acode{binder}   & $\tvar,\; (A\OPandSYMB B)^{\tvar} \vdash B\OPandSYMB A^{\tvar}$          & 2 &  5 &  5 & 354 & 30 \\
\midrule
$\RULEnextR$    & $\Gamma\vdash t\mathrel{\OPtprevSYMB} t_1,\ \Gamma\vdash A^{t_1}\ /\ \Gamma\vdash\OPnextSYMB A^{t}$ & 1 & 2 & 3 & 142 & 26 \\
$\RULEeventuallyR$    & $\Gamma\vdash t\OPtleSYMB t_1,\ \Gamma\vdash A^{t_1}\ /\ \Gamma\vdash\OPeventuallySYMB A^{t}$ & 2 & 3 & 5 & 164 & 28 \\
$\RULEeventuallybyR$    & $\begin{array}[t]{@{}l@{}}\Gamma\vdash t\OPtleSYMB t_1,\ \Gamma\vdash t_1\OPtleSYMB t\OPtaddSYMB t_0,\\ \Gamma\vdash A^{t_1}\ /\ \Gamma\vdash\OPeventuallySYMB_{t_0}A^{t}\end{array}$ & 5 & 7 & 10 & 270 & 48 \\
\bottomrule
\end{tabular}
\end{table}

On most goals the search space is small: the search explores fewer than
200 nodes, and the derivations it returns have at most 41 rule
applications.  The exception is \acode{distr}, where the left rule for
disjunction branches twice and the search explores 620 nodes.  It is the
only goal on which Method~1 takes more than 0.12\,s (it takes 248 ms,
more than twice \acode{chain}$_6$'s time).

Method~2 is between 5 and 25 times slower than Method~1, with an
overhead of 0.07 to 1.4\,s per goal. Both methods evaluate the same
search on the same abstract sequent, and both hand Agda the same kind
of term, in which the derivation appears as
$\acode{from-just}\,(\acode{search⊢}\;10\;\Delta\;\dots)$. The
difference comes from the work the macro does before that, at
type-checking time: it inspects the statement through reflection,
infers the type of each subterm to decide whether it is a formula or a
time expression to abstract, builds $\Delta$, $\sigma$ and $\rho$, and
evaluates the search once more to check that it succeeds, in order to
report an error otherwise. The first part grows with the size of the
statement, as the chains show, and the last one with the cost of the
search, as \acode{distr} shows. The size of the derivation plays no
role: the derivation found for \acode{→} is almost three times as
large as that for $\RULEeventuallybyR$, yet the macro takes about the
same time on both.

\acode{plug-⊢} takes a derivation as argument, and how the macro writes
this argument matters.  As in Method~1, it writes it as the expression
\[
  \acode{from-just}\,(\acode{search⊢}\;10\;\Delta\;\dots),
\]
i.e., as the computation that produces the derivation, and not as the
derivation written out in full, i.e., an explicit term built from the
constructors of $\der$ (such as \acode{ax}, $\RULEimpR$ and
${\RULEimpL}$). Both denote the same derivation. For $\varphi
\OPimpSYMB \varphi$, the explicit form is $\acode{→I}\;\acode{ax}$
together with all the implicit arguments that record the context,
annotation and formulas at each step; this is small, but for larger
goals it is not (see below). For the expression, as explained in
Sec.~\ref{sec:automation:subst}, Agda evaluates the search until it
returns a $\Just$, but does not type-check the derivation it contains,
and does not run \acode{plug-⊢}.
An explicit derivation, in contrast, is handed to Agda as a term to
type-check, exactly as if it had been typed into the file, so it must
be type-checked like a proof written by hand, rule application by rule
application, including the implicit arguments that record the context
and the formulas at every step. This is costly for large derivations.
For example, consider the goal \acode{→} of
Table~\ref{tab:automation:experiments}, i.e., $\vdash (A\OPimpSYMB
B)\OPimpSYMB(B\OPimpSYMB C)\OPimpSYMB A\OPimpSYMB C^{t}$. With fuel
10, the derivation found for it contains a chain of redundant
$\RULEimpL$ steps. After three applications of $\RULEimpR$, the goal
is $C$, and the search applies $\RULEimpL$ to the hypothesis
$B\OPimpSYMB C$, which requires a proof of $B$. To prove $B$, the
depth-first search again first tries $\RULEimpL$ on $B\OPimpSYMB C$,
which requires $B$ once more, and so on until the fuel runs out, at
which point it uses $A\OPimpSYMB B$ instead. Each of these nested
attempts then succeeds, so all of them end up in the derivation. This
derivation has 28 rule applications (against 8 for the smallest fuel
that works), and its quoted normal form has about 9{,}000 nodes.
Re-checking it takes about 2\,s, against 0.44\,s for the whole macro
when it gives the \acode{from-just} expression instead.

The chains \acode{→} (i.e., \acode{chain}$_2$) to \acode{chain}$_6$
show how the cost grows with the size of the goal. The fuel needed
grows with the length of the chain, from 5 to 10, the number of
explored nodes from 29 to 175, the time of Method~1 from 28 to
82\,ms, and that of Method~2 from 0.27 up to 1.41\,s.
The derivation size, in contrast, does not grow steadily: it is 41 for
\acode{chain}$_3$ but only 20 for \acode{chain}$_6$. This is the due
to the redundant $\RULEimpL$ steps described above for \acode{→}.
Sizes are measured with fuel 10, and the less fuel a goal needs, the
more fuel is left for the depth-first search to spend on such detours
before it reaches the right hypothesis. With the smallest fuel that
works, the sizes are 11, 14, 17 and 20 for \acode{chain}$_3$ to
\acode{chain}$_6$, i.e., three rule applications per extra link.
Proofs become long-running when the search fails, either because the
goal is not provable or because the fuel is too small. The search then
explores the whole search tree up to the fuel, whose size grows
exponentially with the fuel because of the branching left rules. For
instance, \acode{chain}$_7$ is not solved with fuel 10: the search
fails after exploring 1{,}049, 3{,}829 and 15{,}197 nodes with fuel 8,
9 and 10, which take 0.1, 0.4 and 1.6\,s, respectively. This cost lies
in the search, so it affects both methods alike. The whole set of
examples (see \myurl{\VGITe}) type-checks in about 52\,s, of which
about 10\,s is spent loading the development.

Beyond the examples above, the automation tool also proves
Lem.~\ref{lem:forwarding-property}~\flinka{84}, the main property of
the forwarding example from Sec.~\ref{sec:application-simple}. Since
its proof uses~$\RULEalwaysL$, which requires classical reasoning, we
extended $\acode{search⊢}$ with an optional parameter standing for the
law of excluded middle, which enables the classical~$\RULEalwaysL$
rule during search.

\section{Conclusion}
\label{sec:conclusion}

We have introduced \calc, a calculus for verifying time-related
properties of fault-tolerant distributed systems, extending TPTL +
Past with quantifiers and a general comparison operator,
featuring a sequent calculus that relies on intervals to precisely
capture timing constraints.
\calc allows formalizing and verifying non-trivial properties of
fault-tolerant algorithms, including one of Pistis' key properties.
We now plan on exploring a few directions.

Based on \calc, we plan on designing a type system tailored to the
unique demands of distributed systems, that would not only allow
modelling their properties, but also allow reasoning about running
code.

Another direction is to build on the architecture of PRISM to
adapt existing model checkers and automated verification techniques
to \calc.
While subsets of \calc could be model checked by PRISM, those
techniques will need to be adapted to handle \calc's freeze and
distributed operators.

Finally, in addition to the examples presented above, we have
specified two other problems in \calc: leader election in a ring and
Byzantine consistent broadcast. Both are presented in
Appx.~\ref{sec:further-examples} and we leave it for future work to
automate their proofs in Agda using the automation tool presented in
Sec.~\ref{sec:automation}.

\section*{AI Disclosure}

While the Agda formalisation was directed and reviewed by the authors,
AI was used to assist with the development of some of the
formalisation. AI was also used to correct typographical errors and to
assist in drafting the sections that discuss the formalisation. All
such text was reviewed and edited by the authors.

\printbibliography

\appendix

\section{Further Examples}
\label{sec:further-examples}

In addition to the forwarding (FW) example provided in
Sec.~\ref{sec:application-simple}, we further illustrate here aspects
of \calc through two other simple examples: leader election in a ring
(LER) in Sec.~\ref{sec:leader-election}, and Byzantine consistent
broadcast (BCB) in Sec.~\ref{sec:bcb}.
All three examples illustrate how to capture message passing patterns
in distributed systems, where agents exchange information by
sending/receiving messages.
Furthermore, while the LER and BCB examples involve a logical,
qualitative notion of time, the FW example involves a physical,
quantitative notion of time thanks to formulas of the form
$\OPeventuallyby{t}{\varphi}$.
Finally, the LER example involves a typical inductively proven
property, while BCB involves some typical pattern,
where agents can only make progress when they have received
enough messages from the other agents in the system.

\subsection{Leader Election in a Ring}
\label{sec:leader-election}

\begin{figure}[!t]
\begin{center}
  \includegraphics[width=0.35\textwidth]{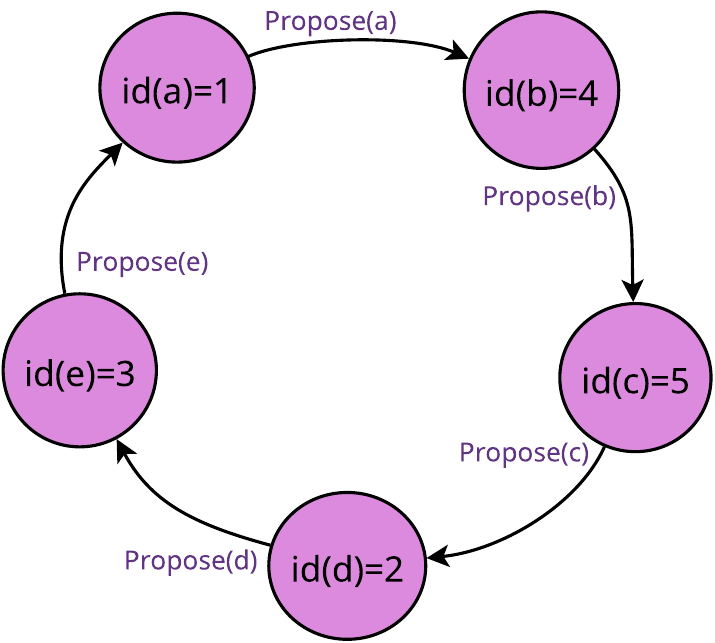}
\end{center}
\caption{Leader Election in a Ring}
\label{fig:leader-election-in-a-ring}
\end{figure}

The leader election in a ring problem involves selecting a single
leader from a group of agents organized in a ring.
Each agent has a unique identifier (id), and the goal is to elect the agent
with the highest id.
Agents do not have knowledge of the entire ring, and messages can only
be passed to the adjacent node according to a fixed (e.g., clockwise)
direction.
Leader election is crucial for coordination in distributed systems,
such as for database replication or time synchronization.

A key property that the algorithm must fulfill is
\emph{liveness}, which guarantees that all agents eventually receive the outcome
of the election, i.e., the identity of the leader:
  \begin{equation}
     \OPall{a}{\LtyAgent}{
     \OPex{b,c}{\LtyAgent}{
       \OPat{\OPtzero}{\OPeventually{\ATreceive{a}{\mathtt{Elected}(c)}{b}}}}
     }
     \label{equ:ring-election-liveliness}
   \end{equation}
The $\OPatSYMB$ operator is used to state that a formula is
true at a specific point in time:
$\OPat{t}{\varphi}\DEF\OPfreeze{x}{(\OPand{\varphi}{\OPteq{x}{t}})}$.
Both $\election{a}$ and $\elected{a}$ are pieces of data
capturing the fact that $a$ was proposed to be the leader, and elected
to be the leader, respectively.

We now present a formalization of the classical ring election
algorithm~\cite{Lann:1977:ifip,Chang+Roberts:1979:cacm} in \calc.
We assume no failures, a synchronous system, a binary predicate $>$
on agents, where $a>b$ captures the fact that $a$ has a higher
identifier than~$b$, and a function $\nhbrSYMB:\LtyAgent\to\LtyAgent$
returning the neighbour of each agent in the ring.

\emph{\textbf{Initialisation}}: Each agent $a$ suggests itself as the leader, sending its id in a
        proposal to its neighbour (see Fig.~\ref{fig:leader-election-in-a-ring}):%
\begin{equation}
   \OPall{a}{\LtyAgent}{
        \OPat{\OPtzero}{
          \ATsend{a}{\election{a}}{[\nhbr{a}]}
   }}
   \label{equ:ring-election-initiation}
\end{equation}

\emph{\textbf{Proposal Propagation}}: When an agent~$a$ receives
        a $\election{c}$ message, it compares $c$'s id to its own id.
 If $c$'s id is greater than $a$'s, $a$ simply forwards the message to its
 neighbour, and if $c$'s id is smaller than  $a$'s, $a$ forwards its own
 id to its neighbour:
\begin{equation}
    \OPall{a,b,c}{\LtyAgent}{
        \OPalways{\lpar\!\!\!\OPimpb
        {\;\ATreceive{a}{\election{c}}{b}}
        {\hspace*{-1in}\OPandb
         {(\OPimp
          {c>a}
          {\ATsend{a}{\election{c}}{[\nhbr{a}]}})}
         {(\OPimp
          {a>c}
          {\ATsend{a}{\election{a}}{[\nhbr{a}]}\rpar})}}}}
    \label{equ:ring-election-greater-propagation}
\end{equation}

\emph{\textbf{Leader Determination}}: If $a$ receives a proposal containing its own id, it
         means that $a$ has the highest id. It then sends
         an $\elected{a}$ message around the ring to announce its
         election as leader.
\begin{equation}
    \OPall{a,b}{\LtyAgent}{
        \OPalways{\lpar\!\!\!\OPimp{\;\;\ATreceive{a}{\election{a}}{b}}
               {\ATsend{a}{\elected{a}}{[\nhbr{a}]}\rpar}}}
    \label{equ:ring-election-election}
\end{equation}

\emph{\textbf{Leader Propagation}}: When an agent $a$ receives an
         $\elected{b}$ message, it knows that $b$ is the leader, and forwards the message
         to its neighbour (unless $a$ is $b$):
\begin{equation}
    \OPallb{a,b,c}{\LtyAgent}{\hspace*{0.1in}
        \OPalways{\lpar\!\!\!\OPimp
        {\;\;\ATreceive{a}{\elected{b}}{c}}
        {\OPimp{b>a}{\ATsend{a}{\elected{b}}{[\nhbr{a}]}\rpar}}}}
    \label{equ:ring-election-dissemination}
\end{equation}

The algorithm ends when the leader has received an $\electedSYMB$
message with its own id.

Liveness (election terminates) is proved by induction on
time
using the following invariant, which states that at time~$x$, if $a$
has not yet received a message informing it that it is the leader (the
$\mathtt{leader}$ operator), then it must be receiving the highest id
among its $x$ predecessors in the ring (the $\mathtt{rcvMaxId}$
operator), where $\mathtt{max}(a,x)$ is the agent with the highest id
among the $x$ predecessors of $a$ in the ring (it returns $a$ when
$x$ is $0$):
\begin{equation}
    \OPall{a}{\LtyAgent}{
    \OPnext{\OPalways{
    (\OPfreeze{x}{
        \OPimp{\OPneg{\mathtt{leader}(a)}}
                {\mathtt{rcvMaxId}(a,\mathtt{max}(a,x))}})
    }}}
\label{equ:ring-election-invariant}
\end{equation}
where
$$\begin{array}{lll}
\mathtt{leader}(a) &\!\! \DEF\!\! & \OPex{c}{\LtyAgent}{\OPbefore{\ATreceive{a}{\election{a}}{c}}}
\\
\mathtt{rcvMaxId}(a,b) &\!\! \DEF\!\! & \OPex{c}{\LtyAgent}{\ATreceive{a}{\election{b}}{c}}
\end{array}$$
This invariant allows concluding that the election phase terminates
after $N$ steps, where $N$ is the number of agents in the ring, with
all agents having received the highest agent/leader, including the
leader itself.
Eq.~\eqref{equ:ring-election-election} can then be used for the leader
to start the final phase and eq.~\eqref{equ:ring-election-dissemination}
guarantees that the $\electedSYMB$ message will go around the ring
back to the leader marking the end of the execution of the algorithm.

We further consider that the agents advance in lockstep, meaning that
each message takes exactly one unit of time to reach its destination:
\begin{equation}
    \OPall{a,b}{\LtyAgent}{
    \OPall{d}{\LtyData}{
        \OPalways{\lpar\OPimp{\ATsend{a}{d}{[b]}}{\OPnext{\ATreceive{b}{d}{a}}}\rpar}
    }}
\label{equ:ring-election-lockstep}
\end{equation}

Let us now sketch the proof of the above invariant.
In the base case where the time $x=0$, $\mathtt{max}(a,x)=a$,
which satisfies the invariant, because each agent is assumed to have
received a proposal for itself at initialisation as captured by
eq.~\eqref{equ:ring-election-initiation}.

For the induction step, assume the property holds for time~$x$. Then,
at time $x+1$, the predecessor of agent $a$, say~$b$, if not yet the
leader, has received
$\mathtt{max}(b,x)$, which is the agent with the largest id among
the~$x$ predecessors of~$b$.
When $b$ receives this information, it compares its own id with
$\mathtt{max}(b,x)$. If $b>\mathtt{max}(b,x)$ or $b<\mathtt{max}(b,x)$, we apply
eq.~\eqref{equ:ring-election-greater-propagation}, sending
$\election{b}$ or
$\election{\mathtt{max}(b,x)}$ to~$a$, respectively.
In both cases $a$ receives $\mathtt{max}(a,x+1)$ at time $x+1$ using
eq.~\eqref{equ:ring-election-lockstep}.

\subsection{Byzantine Consistent Broadcast}
\label{sec:bcb}

\begin{figure}[!t]
\begin{center}
  \includegraphics[width=0.35\textwidth]{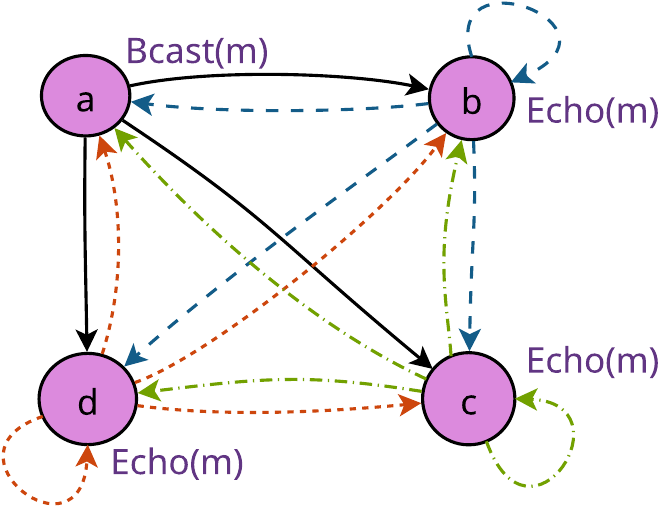}
\end{center}
\caption{Byzantine Consistent Broadcast}
\label{fig:byzantine-consistent-broadcast}
\end{figure}

The Byzantine Consistent
Broadcast~\cite[Sec.~3.10]{cachin_introduction_2011} (BCB) is a
fundamental broadcast primitive designed for the Byzantine fault
model, where faulty agents arbitrarily deviate from their specification, even
deliberately trying to prevent an algorithm from achieving its goals.
It requires $3f+1$ agents to tolerate up to $f$ faulty agents.

Every BCB instance has a unique designated sender initiating the
broadcast of a message. BCB guarantees the \emph{validity} property,
stating that if a correct agent $a$
broadcasts a message $m$, then every correct agent eventually
delivers $m$. This property is stated as follows in \calc, where
$\broadcast{m}, \Echo{m}, \Deliver{m}\in\LtyData$ capture the
broadcast, echoing, and delivery of a message~$m$ (also a piece of
data), respectively:
\begin{equation}
    \OPall{a,b}{\LtyAgent}{
    \OPall{m}{\LtyData}{
    \OPimp
        {\ATcorrect{a}}
        {\OPimpb
          {\ATcorrect{b}}
          {\hspace*{-2.1in}\OPalways{\lpar\OPimp
             {\ATinternal{a}{\broadcast{m}}}
             {\OPeventually{\ATinternal{b}{\Deliver{m}}}\rpar}}
          }
        }
    }}
\label{equ:bcb-validity}
\end{equation}

\emph{\textbf{Broadcast}}:
  An agent initiates a broadcast through an internal broadcast event of
   the form $\ATinternal{a}{\broadcast{m}}$. When
   initiating a broadcast, the agent sends a broadcast message to all
   other agents, where $N$ is the collection of agents in the system:
  \begin{equation}
   \OPall{a}{\LtyAgent}{\OPall{m}{\LtyData}{\OPimpb
       {\ATcorrect{a}}
       {
         \hspace*{-1.0in}
         \OPalways{\lpar\OPimp{\ATinternal{a}{\broadcast{m}}}{\ATsend{a}{\broadcast{m}}{N}\rpar}}}
     }
   }
   \label{equ:bcb-validity-bcast}
  \end{equation}

\emph{\textbf{Echo}}:  When an agent receives a broadcast message, it then sends an echo message to
  all other agents:
  \begin{equation}
   \OPall
     {a,b}{\LtyAgent}
        {\OPall
           {m}{\LtyData}{\OPimpb
             {\ATcorrect{a}}
             {\hspace*{-1.0in}\OPalways{\lpar\OPimp{\ATreceive{a}{\broadcast{m}}{b}}{\ATsend{a}{\Echo{m}}{N}\rpar}}}
           }
        }
     \label{equ:bcb-validity-echo}
  \end{equation}

\emph{\textbf{Deliver}}:
  Agents deliver on receipt of $2f{+}1$ echoes:
  \begin{equation}
    \OPall{a}{\LtyAgent}{\OPall{m}{\LtyData}{
        \OPalways{\lpar\OPimpb
        {\ATcorrect{a}}
        {\hspace*{-0.7in}\OPimpb
          {\left(\OPexbb{B}{\LtyAgents}
              {\hspace*{0.05in}
               \OPandb
                {\OPsize{B}{2f+1}}
                {\OPall{b}{\LtyAgent}{
                   \OPimpb
                   {\OPin{b}{B}}
                   {\hspace*{-0.6in}\OPeventually{\ATreceive{a}{\Echo{m}}{b}}}}}}\right)}
          {\OPeventually{\ATinternal{a}{\Deliver{m}}}\rpar}
        }}}
    }
     \label{equ:bcb-validity-deliver}
  \end{equation}

\emph{\textbf{Reliable Communication}}:
   We assume reliable communication, i.e., any message sent by a
   correct agent will eventually be received by all other correct agents:
  \begin{equation}
  \OPall{a}{\LtyAgent}{\OPall{d}{\LtyData}{\OPallb{B}{\LtyAgents}{\hspace*{-1.2in}
    \OPalways{\lpar\OPimp
      {\ATcorrect{a}}
      {\OPimpb
        {\ATsend{a}{d}{B}}
        {\hspace*{-0.8in}\OPall{b}{\LtyAgent}
          {\OPimpb{\OPin{b}{B}}{\hspace*{-0.6in}\OPimp{\ATcorrect{b}}{\OPeventually{\ATreceive{b}{d}{a}}\rpar}}}}}}}}}
     \label{equ:bcb-validity-reliable}
  \end{equation}

\emph{\textbf{Fault Threshold}}:
  Finally, we assume that there are at least $2f+1$ correct agents out
  of $3f+1$, where $P$ below is a predicate on agents:
  \begin{equation}
  \OPalways{\lpar\!\!\!\OPimpb
  {(\OPall{b}{\LtyAgent}{\OPimp{\ATcorrect{b}}{P(b)}})}
  {\hspace*{0.0in}\OPex{B}{\LtyAgents}{
        \OPandbb
        {\OPsize{B}{2f+1}}
        {\hspace*{-0.6in}\OPall{b}{\LtyAgent}{\OPimp{\OPin{b}{B}}{\OPand{\ATcorrect{b}}{P(b)}\rpar}}}}}}
  \label{equ:2f+1-correct-nodes}
  \end{equation}

Fig.~\ref{fig:byzantine-consistent-broadcast} shows a BCB
execution, in a system with $3f+1=4$ agents, with at most $f=1$
Byzantine agent. Agent~$a$ broadcasts message~$m$, which is then
echoed by all agents (dashed blue lines for~$b$, dashed and dotted
green lines for~$c$, and dotted orange lines for $d$). After receiving
$2f+1=3$ echoes from $b,c,d$, agent~$b$, for example, delivers.

Let us now sketch a proof of the validity property. Assume correct
agents $a$ and $b$ such that $a$ initiates a broadcast of~$m$, i.e.,
$\ATinternal{a}{\broadcast{m}}$ holds.
We need to prove that $b$ eventually delivers $m$, i.e.,
$\OPeventually{\ATinternal{b}{\Deliver{m}}}$.
By eq.~\eqref{equ:bcb-validity-bcast}, $a$ also sends $\broadcast{m}$ to
all agents in $N$.
By eq.~\eqref{equ:bcb-validity-reliable}, all correct agents in $N$
receive $\broadcast{m}$ and by eq.~\eqref{equ:bcb-validity-echo} they
send $\Echo{m}$ messages to all other agents in $N$, including~$b$.
Therefore, by eq.~\eqref{equ:bcb-validity-reliable}, $b$ will
receive $\Echo{m}$ from all correct agents.
Finally, by eq.~\eqref{equ:bcb-validity-deliver} and
eq.~\eqref{equ:2f+1-correct-nodes}, agent $b$ delivers $m$.

\section{Full Set of Rules}
\label{sec:full-set-of-rules}

\paragraph{Propositional Logic}
\calc's propositional logic rules are as follows:
\begin{small}
\[
\begin{array}{c}
  \inference{\nseq{\Gamma,\Delta}{\annot_1}}
            {\seq{\Gamma,\hyp{\OPfalse}{\annot_1},\Delta}{\annot}{\varphi}}
            [$\RULEfalseL$]
  \hspace{0.2in}
  \inference{\seq{\Gamma}{\annot}{\varphi} & \seq{\Gamma}{\annot}{\psi}}
            {\seq{\Gamma}{\annot}{\OPand{\varphi}{\psi}}}
            [$\RULEandR$]
  \hspace{0.2in}
  \inference{\seq{\Gamma,\hyp{\varphi}{\annot},\hyp{\psi}{\annot},\Delta}{\annot_1}{\delta}}
            {\seq{\Gamma,\hyp{\OPand{\varphi}{\psi}}{\annot},\Delta}{\annot_1}{\delta}}
            [$\RULEandL$]
\end{array}
\]
\vspace{-0.05in}
\[
\begin{array}{c}
  \inference{\seq{\Gamma}{\annot}{\varphi}}
            {\seq{\Gamma}{\annot}{\OPor{\varphi}{\psi}}}
            [$\RULEorRa$]
  \hspace{0.2in}
  \inference{\seq{\Gamma}{\annot}{\psi}}
            {\seq{\Gamma}{\annot}{\OPor{\varphi}{\psi}}}
            [$\RULEorRb$]
  \hspace{0.2in}
  \inference{\seq{\Gamma,\hyp{\varphi}{t},\Delta}{\annot}{\delta} & \seq{\Gamma,\hyp{\psi}{t},\Delta}{\annot}{\delta}}
            {\seq{\Gamma,\hyp{\OPor{\varphi}{\psi}}{t},\Delta}{\annot}{\delta}}
            [$\RULEorL$]
\end{array}
\]
\vspace{-0.05in}
\[
\begin{array}{c}
  \inference{}
            {\seq{\Gamma}{\annot}{\OPtrue}}
            [$\RULEtrueR$]
  \hspace{0.2in}
  \inference{\seq{\Gamma,\hyp{\varphi}{t}}{t}{\psi}}
            {\seq{\Gamma}{t}{\OPimp{\varphi}{\psi}}}
            [$\RULEimpR$]
  \hspace{0.2in}
  \inference{\seq{\Gamma,\Delta}{t}{\varphi} & \seq{\Gamma,\hyp{\psi}{t},\Delta}{\annot}{\delta}}
            {\seq{\Gamma,\hyp{\OPimp{\varphi}{\psi}}{t},\Delta}{\annot}{\delta}}
            [$\RULEimpL$]
\end{array}
\]
\end{small}

\calc also includes the following rules. We often make silent use of
the $\RULEcut$ rule to duplicate a hypothesis whenever using a
``left'' (or elimination) rule on that hypothesis if we wish to retain
it, because \calc's ``left'' rules discard the formulas being
eliminated.
\begin{small}
\[
\begin{array}{l@{\hspace{0.2in}}l}
  \inference{\seq{\Gamma,\Delta}{\annot}{\varphi}}
            {\seq{\Gamma,\hyp{\psi}{\annot_1},\Delta}{\annot}{\varphi}}
            [$\RULEthin$]
  &
  \inference{\seq{\Gamma}{\annot_1}{\psi} & \seq{\Gamma,\hyp{\psi}{\annot_1}}{\annot}{\varphi}}
            {\seq{\Gamma}{\annot}{\varphi}}
            [$\RULEcut$]
\end{array}
\]
\vspace{-0.05in}
\[
\begin{array}{l@{\hspace{0.2in}}l}
  \inference{}
            {\seq{\Gamma,\hyp{\varphi}{\annot}}{\annot}{\varphi}}
            [$\RULEvar$]
  &
  \inference{}
            {\seq{\Gamma}{\annot}{\OPin{i}{[\dots,i,\dots]}}}
            [$\RULEinR$]
\end{array}
\]
\end{small}

\paragraph{Non-Empty Annotations}
\begin{small}
\[
\begin{array}{l@{\hspace{0.1in}}l@{\hspace{0.1in}}l}
\inference{}
          {\nseq{\Gamma}{\ANno}}
          []
&
\inference{\seqb{\Gamma}{\OPtle{t_1}{t_2}}}
          {\nseq{\Gamma}{\ANintcc{t_1}{t_2}}}
          []
&
\inference{\seqb{\Gamma}{\OPtlt{t_1}{t_2}}}
          {\nseq{\Gamma}{\ANintco{t_1}{t_2}}}
          []
\\ \\
\inference{}
          {\nseq{\Gamma}{t}}
          []
&
\inference{\seqb{\Gamma}{\OPtlt{t_1}{t_2}}}
          {\nseq{\Gamma}{\ANintoc{t_1}{t_2}}}
          []
&
\inference{\seqb{\Gamma}{\OPtlt{t_1}{t}} & \seqb{\Gamma}{\OPtlt{t}{t_2}}}
          {\nseq{\Gamma}{\ANintoo{t_1}{t_2}}}
          []
\end{array}
\]
\end{small}

\paragraph{Temporal Logic Rules}
In the following rules $\tvar$ is a ``fresh'' variable:
\begin{small}
\[
\begin{array}{c@{\hspace{0.1in}}c}
  \inference{\seqb{\Gamma}{\OPtprev{t}{t_1}}
             & \seq{\Gamma}{t_1}{\varphi}}
            {\seq{\Gamma}{t}{\OPnext{\varphi}}}
            [$\RULEnextR$]
  &
    \inference{\seqb{\Gamma}{\OPtle{t}{t_1}}
               & \seq{\Gamma}{t_1}{\psi}
               & \seq{\Gamma}{\ANintco{t}{t_1}}{\varphi}}
              {\seq{\Gamma}{t}{\OPuntil{\varphi}{\psi}}}
              [$\RULEuntilR$]
\end{array}
\]
\vspace{-0.05in}
\[
\begin{array}{c@{\hspace{0.1in}}c}
  \inference{\seq{\Gamma,\OPtprev{t}{\tvar},\hyp{\varphi}{\tvar},\Delta}{\annot}{\delta}}
            {\seq{\Gamma,\hyp{\OPnext{\varphi}}{t},\Delta}{\annot}{\delta}}
            [$\RULEnextL$]
  &
    \inference{\seq{\Gamma,\OPtle{t}{\tvar},\hyp{\psi}{\tvar},\hyp{\varphi}{\ANintco{t}{\tvar}},\Delta}{\annot}{\delta}}
              {\seq{\Gamma,\hyp{\OPuntil{\varphi}{\psi}}{t},\Delta}{\annot}{\delta}}
              [$\RULEuntilL$]
\end{array}
\]
\end{small}

\begin{small}
\[
\begin{array}{c@{\hspace{0.1in}}c}
  \inference{\seqb{\Gamma}{\OPtprev{t_1}{t}}
             & \seq{\Gamma}{t_1}{\varphi}}
            {\seq{\Gamma}{t}{\OPprev{\varphi}}}
            [$\RULEprevR$]
  &
  \inference{\seqb{\Gamma}{\OPtle{t_1}{t}}
             & \seq{\Gamma}{t_1}{\psi}
             & \seq{\Gamma}{\ANintoc{t_1}{t}}{\varphi}}
            {\seq{\Gamma}{t}{\OPsince{\varphi}{\psi}}}
            [$\RULEsinceR$]
\end{array}
\]
\vspace{-0.05in}
\[
\begin{array}{c@{\hspace{0.1in}}c}
  \inference{\seq{\Gamma,\OPtprev{\tvar}{t},\hyp{\varphi}{\tvar},\Delta}{\annot}{\delta}}
            {\seq{\Gamma,\hyp{\OPprev{\varphi}}{t},\Delta}{\annot}{\delta}}
            [$\RULEprevL$]
  &
  \inference{\seq{\Gamma,\OPtle{\tvar}{t},\hyp{\psi}{\tvar},\hyp{\varphi}{\ANintoc{\tvar}{t}},\Delta}{\annot}{\delta}}
            {\seq{\Gamma,\hyp{\OPsince{\varphi}{\psi}}{t},\Delta}{\annot}{\delta}}
            [$\RULEsinceL$]
\end{array}
\]
\end{small}

\paragraph{Timed Rules}
The rules for the comparison and freeze operators are shown below. We
only show some of the comparison rules, as they reflect the axioms
in Fig.~\ref{fig:world-axioms}.
Furthermore, \calc's $\RULEiccL$, $\RULEiocL$, $\RULEicoL$,
$\RULEiooL$, $\RULEiccR$, $\RULEiocR$, $\RULEicoR$, and $\RULEiooR$
rules allow instantiating intervals, some of which are shown below
(the others are of the same form).
\begin{small}
\[
\begin{array}{c}
  \inference{}
            {\seq{\Gamma,\hyp{\OPtle{t_1}{t_2}}{},\Delta}{\annot}{\OPtle{t_1}{t_2}}}
            [$\RULEvarle$]
  \hspace{0.1in}
  \inference{\seq{\Gamma}{t}{\Msub{\varphi}{\tvar}{t}}}
            {\seq{\Gamma}{t}{\OPfreeze{\tvar}{\varphi}}}
            [$\RULEfreezeR$]
\end{array}
\]
\vspace{-0.05in}
\[
\begin{array}{c}
  \inference{}
            {\seq{\Gamma}{\annot}{\OPteq{\OPtadd{t_1}{t_2}}{\OPtadd{t_2}{t_1}}}}
            [$\RULEaddSym$]
  \hspace{0.1in}
  \inference{\seq{\Gamma,\hyp{\Msub{\varphi}{\tvar}{t}}{t},\Delta}{\annot}{\psi}}
            {\seq{\Gamma,\hyp{\OPfreeze{\tvar}{\varphi}}{t},\Delta}{\annot}{\psi}}
            [$\RULEfreezeL$]
\end{array}
\]
\end{small}

\begin{small}
\[
\begin{array}{c}
\inference{\seq{\Gamma,\hyp{\varphi}{t},\Delta}{\annot}{\psi}
           & \seqb{\Gamma,\Delta}{\OPtle{t_1}{t}}
           & \seqb{\Gamma,\Delta}{\OPtle{t}{t_2}}}
          {\seq{\Gamma,\hyp{\varphi}{\ANintcc{t_1}{t_2}},\Delta}{\annot}{\psi}}
          [$\RULEiooL$]
\\ \\
\inference{\seq{\Gamma,\OPtle{t_1}{\tvar},\OPtle{\tvar}{t_2}}{\tvar}{\phi}}
          {\seq{\Gamma}{\ANintcc{t_1}{t_2}}{\phi}}
          [$\RULEiooR$]
\end{array}
\]
\end{small}

\paragraph{Predicate Logic Rules}
The predicate rules are as follows, where $v'$ is a ``fresh'' variable
and $X$ is a term of type $T$:
\begin{small}
\[
\begin{array}{ll}
  \inference{\seq{\Gamma}{\annot}{\Msub{\varphi}{v}{v'}}}
            {\seq{\Gamma}{\annot}{\OPall{v}{T}{\varphi}}}
            [$\RULEforallR$]
  &
  \inference{\seq{\Gamma,\hyp{\Msub{\varphi}{v}{X}}{\annot_1},\Delta}{\annot}{\psi}}
            {\seq{\Gamma,\hyp{\OPall{v}{T}{\varphi}}{\annot_1},\Delta}{\annot}{\psi}}
            [$\RULEforallL$]
\end{array}
\]
\vspace{-0.05in}
\[
\begin{array}{ll}
  \inference{\seq{\Gamma}{\annot}{\Msub{\varphi}{v}{X}}}
            {\seq{\Gamma}{\annot}{\OPex{v}{T}{\varphi}}}
            [$\RULEexistsR$]
  &
  \inference{\seq{\Gamma,\hyp{\Msub{\varphi}{v}{v'}}{t},\Delta}{\annot}{\psi}}
            {\seq{\Gamma,\hyp{\OPex{v}{T}{\varphi}}{t},\Delta}{\annot}{\psi}}
            [$\RULEexistsL$]
\end{array}
\]
\end{small}

\paragraph{Induction}
The induction rule is as follows, where $\tvar$ does not occur free in
$\Gamma$:
\begin{small}
\[
\inference{\seq{\Gamma,\hyp{\OPfreeze{\tvar}{\varphi}}{\ANintco{\OPtzero}{\tvar}}}{\tvar}{\varphi}}
          {\seq{\Gamma}{\tvar}{\varphi}}
          [$\RULEind$]
\]
\end{small}

\paragraph{Classical Rules}
\begin{small}
\[
\inference{}
          {\seq{\Gamma}{\annot}{\OPor{\varphi}{\OPneg{\varphi}}}}
          [$\RULElem$]
\]
\end{small}

\paragraph{Derived Rules}
In all the rules below,~$\tvar$ stands for a ``fresh'' variable.
\begin{small}
\[
\begin{array}{l@{\hspace{0.1in}}l}
  \inference{\seq{\Gamma}{t_1}{\varphi} & \seqb{\Gamma}{\OPtle{t}{t_1}}}
            {\seq{\Gamma}{t}{\OPeventually{\varphi}}}
            [$\RULEeventuallyR$]
  &
  \inference{\seq{\Gamma,\hyp{\varphi}{\tvar},\OPtle{t_1}{\tvar},\Delta}{t}{\psi}}
            {\seq{\Gamma,\hyp{\OPeventually{\varphi}}{t_1},\Delta}{t}{\psi}}
            [$\RULEeventuallyL$]
\end{array}
\]
\end{small}

\begin{small}
\[
\begin{array}{l@{\hspace{0.1in}}l}
  \inference{\seq{\Gamma,\OPtle{t}{\tvar}}{\tvar}{\varphi}}
            {\seq{\Gamma}{t}{\OPalways{\varphi}}}
            [$\RULEalwaysR$]
  &
  \inference{\seq{\Gamma,\hyp{\varphi}{t_2},\Delta}{t}{\psi}
             & \seqb{\Gamma,\Delta}{\OPtle{t_1}{t_2}}}
            {\seq{\Gamma,\hyp{\OPalways{\varphi}}{t_1},\Delta}{t}{\psi}}
            [$\RULEalwaysL$]
\end{array}
\]
\end{small}

\begin{small}
\[
\begin{array}{l}
  \inference{\seq{\Gamma}{t_1}{\varphi} & \seqb{\Gamma}{\OPtle{t}{t_1}} & \seqb{\Gamma}{\OPtle{t_1}{\OPtadd{t_0}{t}}}}
            {\seq{\Gamma}{t}{\OPeventuallyby{t_0}{\varphi}}}
            [$\RULEeventuallybyR$]
  \\ \\
  \inference{\seq{\Gamma,\hyp{\varphi}{\tvar},\OPtle{t_1}{\tvar},\OPtle{\tvar}{\OPtadd{t_0}{t_1}},\Delta}{t}{\psi}}
            {\seq{\Gamma,\hyp{\OPeventuallyby{t_0}{\varphi}}{t_1},\Delta}{t}{\psi}}
            [$\RULEeventuallybyL$]
\end{array}
\]
\end{small}

\begin{small}
\[
\begin{array}{l}
  \inference{\seq{\Gamma,\OPtle{t}{\tvar},\OPtle{\tvar}{\OPtadd{t_0}{t}}}{t}{\varphi}}
            {\seq{\Gamma}{t}{\OPalwaysby{t_0}{\varphi}}}
            [$\RULEalwaysbyR$]
  \\ \\
  \inference{\seq{\Gamma,\hyp{\varphi}{t_2},\Delta}{t}{\psi}
             & \seqb{\Gamma,\Delta}{\OPtle{t_1}{t_2}}
             & \seqb{\Gamma,\Delta}{\OPtle{t_2}{\OPtadd{t_0}{t_1}}}}
            {\seq{\Gamma,\hyp{\OPalwaysby{t_0}{\varphi}}{t_1},\Delta}{t}{\psi}}
            [$\RULEalwaysbyL$]
\end{array}
\]
\end{small}

\begin{small}
\[
\begin{array}{l@{\hspace{0.1in}}l}
  \inference{\seq{\Gamma}{t_1}{\varphi} & \seqb{\Gamma}{\OPtle{t_1}{t}}}
            {\seq{\Gamma}{t}{\OPbefore{\varphi}}}
            [$\RULEbeforeR$]
  &
  \inference{\seq{\Gamma,\hyp{\varphi}{\tvar},\OPtle{\tvar}{t_1},\Delta}{t}{\psi}}
            {\seq{\Gamma,\hyp{\OPbefore{\varphi}}{t_1},\Delta}{t}{\psi}}
            [$\RULEbeforeL$]
\end{array}
\]
\end{small}

\begin{small}
\[
\begin{array}{l@{\hspace{0.1in}}l}
  \inference{\seq{\Gamma,\OPtle{\tvar}{t}}{\tvar}{\varphi}}
            {\seq{\Gamma}{t}{\OPalwaysbefore{\varphi}}}
            [$\RULEalwaysbeforeR$]
  &
  \inference{\seq{\Gamma,\hyp{\varphi}{t_2},\Delta}{t}{\psi}
             & \seqb{\Gamma,\Delta}{\OPtle{t_2}{t_1}}}
            {\seq{\Gamma,\hyp{\OPalwaysbefore{\varphi}}{t_1},\Delta}{t}{\psi}}
            [$\RULEalwaysbeforeL$]
\end{array}
\]
\end{small}

\begin{small}
\[
\begin{array}{l@{\hspace{0.1in}}l}
  \inference{\seq{\Gamma}{t_1}{\varphi} & \seqb{\Gamma}{\OPtlt{t_1}{t}}}
            {\seq{\Gamma}{t}{\OPstrictbefore{\varphi}}}
            [$\RULEstrictbeforeR$]
  &
  \inference{\seq{\Gamma,\hyp{\varphi}{\tvar},\OPtlt{\tvar}{t_1},\Delta}{t}{\psi}}
            {\seq{\Gamma,\hyp{\OPstrictbefore{\varphi}}{t_1},\Delta}{t}{\psi}}
            [$\RULEstrictbeforeL$]
\end{array}
\]
\end{small}

\section{Proof of eq.~\eqref{equ:simple-example-concl}}
\label{sec:forward-proof}

This section presents a fully worked out proof
of eq.~\eqref{equ:simple-example-concl}, a sketch of which was presented
in Sec.~\ref{sec:examples-of-simple-proofs}.

Let
\[
\begin{array}[t]{lll}
  \Gamma_0 & \DEF & \hypp{eq.~\eqref{equ:simple-example-hyp1}}{t},\hypp{eq.~\eqref{equ:simple-example-hyp2}}{t}
  \\
  \Gamma_1 & \DEF & \Gamma_0,\OPtle{t}{\tvar},\hyp{\ATsend{a}{d}{[b]}}{\tvar},\hyp{\ATreceive{b}{d}{a}}{\tvar'},\OPtle{\tvar}{\tvar'},\OPtle{\tvar'}{\OPtadd{T}{\tvar}}
  \\
  \Gamma_2 & \DEF & \Gamma_1,\hyp{\ATsend{b}{d}{[c]}}{\tvar'}
  \\
  \Gamma_3 & \DEF & \Gamma_2,\hyp{\ATreceive{c}{d}{b}}{\tvar''},\OPtle{\tvar'}{\tvar''},\OPtle{\tvar''}{\OPtadd{T}{\tvar'}}
\end{array}
\]

To prove $\seq{\Gamma_0}{t}{eq.~\eqref{equ:simple-example-concl}}$, we
start with a few introductions to get, in particular, the assumption
that $\hyp{\ATsend{a}{d}{[b]}}{\tvar}$:
\begin{small}
\[
\infer[\RULEforallR (\times{4})]
  {\seq{\Gamma_0}{t}{eq.~\eqref{equ:simple-example-concl}}}
  {\infer[\RULEalwaysR]
    {\seq{\Gamma_0}{t}{\OPalways{(\OPimp{\ATsend{a}{d}{[b]}}{\OPeventuallyby{2T}{\ATreceive{c}{d}{b}}})}}}
    {\infer[\RULEimpR]
      {\seq{\Gamma_0,\OPtle{t}{\tvar}}{\tvar}{\OPimp{\ATsend{a}{d}{[b]}}{\OPeventuallyby{2T}{\ATreceive{c}{d}{b}}}}}
      {\seq{\Gamma_0,\OPtle{t}{\tvar},\hyp{\ATsend{a}{d}{[b]}}{\tvar}}{\tvar}{\OPeventuallyby{2T}{\ATreceive{c}{d}{b}}}}
    }
  }
\]
\end{small}

We now instantiate eq.~\eqref{equ:simple-example-hyp1} to, in
particular, get the assumption $\hyp{\ATreceive{b}{d}{a}}{\tvar'}$:
\begin{small}
\[
\infer[\RULEforallL (\times{4})\ \&\ \RULEalwaysL\ \&\ \RULEvarle]
      {\seq{\Gamma_0,\OPtle{t}{\tvar},\hyp{\ATsend{a}{d}{[b]}}{\tvar}}{\tvar}{\OPeventuallyby{2T}{\ATreceive{c}{d}{b}}}}
      {\infer[\RULEimpL\ \&\ \RULEvar]
        {\seq{\Gamma_0,\OPtle{t}{\tvar},\hyp{\ATsend{a}{d}{[b]}}{\tvar},\hyp{\phi_1}{\tvar}}{\tvar}{\OPeventuallyby{2T}{\ATreceive{c}{d}{b}}}}
        {\infer[\RULEeventuallybyL]
          {\seq{\Gamma_0,\OPtle{t}{\tvar},\hyp{\ATsend{a}{d}{[b]}}{\tvar},\hyp{\phi_2}{\tvar}}{\tvar}{\OPeventuallyby{2T}{\ATreceive{c}{d}{b}}}}
          {\infer[\RULEimpL\ \&\ \RULEinR]
            {\seq{\Gamma_0,\OPtle{t}{\tvar},\hyp{\ATsend{a}{d}{[b]}}{\tvar},\hyp{\phi_3}{\tvar'},\OPtle{\tvar}{\tvar'},\OPtle{\tvar'}{\OPtadd{T}{\tvar}}}{\tvar}{\OPeventuallyby{2T}{\ATreceive{c}{d}{b}}}}
            {\seq{\Gamma_1}{\tvar}{\OPeventuallyby{2T}{\ATreceive{c}{d}{b}}}}
          }
        }
      }
\]
\end{small}
where
$$\begin{array}{lll}
\phi_1 & \DEF & \OPimp{\ATsend{a}{d}{[b]}}{\phi_2}
\\
\phi_2 & \DEF & \OPeventuallyby{T}{(\phi_3)}
\\
\phi_3 & \DEF & \OPimp{\OPin{b}{[b]}}{\ATreceive{b}{d}{a}}
\end{array}$$

We instantiate eq.~\eqref{equ:simple-example-hyp2} to, in
particular, get the assumption $\hyp{\ATsend{b}{d}{[c]}}{\tvar'}$:
\[
\begin{small}
\begin{array}{l}
\infer[\RULEforallL (\times{4})\ \&\ \RULEalwaysL]
      {\seq{\Gamma_1}{\tvar}{\OPeventuallyby{2T}{\ATreceive{c}{d}{b}}}}
      {\infer[\RULEimpL\ \&\ \RULEvar]
        {\seq{\Gamma_1,\hyp{\phi_4}{\tvar'}}{\tvar}{\OPeventuallyby{2T}{\ATreceive{c}{d}{b}}}}
        {\seq{\Gamma_2}{\tvar}{\OPeventuallyby{2T}{\ATreceive{c}{d}{b}}}}
        & \infer[\RULEleTrans\ \&\ \RULEvarle]
        {\seq{\Gamma_1}{\tvar}{\OPtle{t}{\tvar'}}}
        {\hide{\infer[\RULEvarle]
          {\seq{\Gamma_1}{\tvar}{\OPtle{t}{\tvar}}}
          {}
          & \infer[\RULEvarle]
          {\seq{\Gamma_1}{\tvar}{\OPtle{\tvar}{\tvar'}}}
          {}}
        }
      }
\end{array}
\end{small}
\]
where
$$\begin{array}{lll}
\phi_4 & \DEF & \OPimp{\ATreceive{b}{d}{a}}{\ATsend{b}{d}{[c]}}
\end{array}$$

We instantiate eq.~\eqref{equ:simple-example-hyp1} once more to get, in
particular, the assumption $\hyp{\ATreceive{c}{d}{b}}{\tvar''}$:
\[
\begin{small}
\infer[\RULEforallL (\times{4})\ \&\ \RULEalwaysL]
      {\seq{\Gamma_2}{\tvar}{\OPeventuallyby{2T}{\ATreceive{c}{d}{b}}}}
      {\infer[\RULEimpL\ \&\ \RULEvar]
        {\seq{\Gamma_2,\hyp{\phi_5}{\tvar'}}{\tvar}{\OPeventuallyby{2T}{\ATreceive{c}{d}{b}}}}
        {\infer[\RULEeventuallybyL]
          {\seq{\Gamma_2,\hyp{\phi_6}{\tvar'}}{\tvar}{\OPeventuallyby{2T}{\ATreceive{c}{d}{b}}}}
          {\infer[\RULEimpL\ \&\ \RULEinR]
            {\seq{\Gamma_2,\hyp{\phi_7}{\tvar''},\OPtle{\tvar'}{\tvar''},\OPtle{\tvar''}{\OPtadd{T}{\tvar'}}}{\tvar}{\OPeventuallyby{2T}{\ATreceive{c}{d}{b}}}}
            {\seq{\Gamma_3}{\tvar}{\OPeventuallyby{2T}{\ATreceive{c}{d}{b}}}}
          }
        }
        & \infer[\RULEleTrans\ \&\ \RULEvarle]
        {\seq{\Gamma_1}{\tvar}{\OPtle{t}{\tvar'}}}
        {\hide{\infer[\RULEvarle]
          {\seq{\Gamma_1}{\tvar}{\OPtle{t}{\tvar}}}
          {}
          & \infer[\RULEvarle]
          {\seq{\Gamma_1}{\tvar}{\OPtle{\tvar}{\tvar'}}}
          {}}
        }
      }
\end{small}
\]
where
$$\begin{array}{lll}
\phi_5 & \DEF & \OPimp{\ATsend{b}{d}{[c]}}{\phi_6}
\\
\phi_6 & \DEF & \OPeventuallyby{T}{(\phi_7)}
\\
\phi_7 & \DEF & \OPimp{\OPin{c}{[c]}}{\ATreceive{c}{d}{b}}
\end{array}$$

We finally conclude using in particular the derived rule
$\OPeventuallySYMB_{\_}\mbox{R}$, where
$\seq{\Gamma_3}{\tvar}{\OPtle{\tvar''}{\OPtadd{2T}{\tvar}}}$
is trivial but tedious to prove using, among other rules,
$\RULEleTrans$, $\OPtaddSYMB{\mbox{assoc}}$, and $\RULEvarle$:
\[
\begin{small}
\infer[\RULEeventuallybyR]
      {\seq{\Gamma_3}{\tvar}{\OPeventuallyby{2T}{\ATreceive{c}{d}{b}}}}
      {\infer[\RULEleTrans\ \&\ \RULEvarle]
        {\seq{\Gamma_3}{\tvar}{\OPtle{\tvar}{\tvar''}}}
        {\hide{\infer[\RULEvar]
          {\seq{\Gamma_3}{\tvar}{\OPtle{\tvar}{\tvar'}}}
          {}
          & \infer[\RULEvar]
          {\seq{\Gamma_3}{\tvar}{\OPtle{\tvar'}{\tvar'}}}
          {}}
        }
        & \infer[\hide{\RULEleTrans}]
        {\seq{\Gamma_3}{\tvar}{\OPtle{\tvar''}{\OPtadd{2T}{\tvar}}}}
        {\hide{\infer[\RULEvarle]
          {\seq{\Gamma_3}{\tvar}{\OPtle{\tvar''}{\OPtadd{T}{\tvar'}}}}
          {}
          & \seq{\Gamma_3}{\tvar}{\OPtle{\OPtadd{T}{\tvar'}}{\OPtadd{2T}{\tvar}}}
        }}
      }
\end{small}
\]

\end{document}